\documentclass[11pt]{amsart}
\usepackage{amssymb}
\usepackage{amsfonts}
\usepackage{amscd}
\usepackage{amsmath}
\usepackage{mathrsfs}
\usepackage{curves}
\usepackage{epsfig}
\usepackage[margin=1.2in, centering]{geometry}
\usepackage{graphicx}
\usepackage{verbatim}
\usepackage{latexsym}
\usepackage{eucal} 
\usepackage{bm}
\usepackage{color}
\usepackage{algorithm}
\usepackage{algpseudocode}

\numberwithin{equation}{section}
\newtheorem{theorem}{Theorem}[section]
\newtheorem{lemma}[theorem]{Lemma}

\def \mcb {{\mathcal B}}
\def \mcc {{\mathcal C}}

\def \mci {{\mathcal I}}

\def \mcm {{\mathcal M}}

\def \mcr {{\mathcal R}}

\def \mcu {{\mathcal U}}
\def \mcv {{\mathcal V}}

\def \mbr {{\mathbb R}}
\def \mbs {{\mathbb S}}

\def \id {\operatorname{Id}}

\def \supp {\text{supp }}

\def \beqq {\begin{equation}}
\def \eeqq {\end{equation}}
\def \bpf {\begin{proof}}
\def \epf {\end{proof}}
\def \beq {\begin{equation*}}
\def \eeq {\end{equation*}}

\def \eps {\epsilon}

\def \lap {\Delta}
\def \p {\partial}
\def \ha {\frac{1}{2}}

\newcommand{\kron}{\otimes}
\DeclareMathOperator*{\argmin}{arg\,min}                   
\renewcommand{\t} {^{\top}}                                

\renewcommand{\phi}{\mathbf{\varphi}}

\newcommand{\bfone}{{\bf1}}

\newcommand{\bfT}{\mathbf{T}}
\newcommand{\bff}{\mathbf{f}}
\newcommand{\bfA}{\mathbf{A}}

\newcommand{\bfI}{\mathbf{I}}
\newcommand{\bfD}{\mathbf{D}}

\newcommand{\bfb}{\mathbf{b}}

\newcommand{\bfH}{\mathbf{H}}

\newcommand{\bfe}{\mathbf{e}}
\newcommand{\bfu}{\mathbf{u}}

\newcommand{\bfL}{\mathbf{L}}

\newcommand{\bfv}{\mathbf{v}}

\newcommand{\bfS}{\mathbf{S}}

\newcommand{\calN}{\mathcal{N}}

\newcommand{\bfLambda}{{\boldsymbol{\Lambda}}}

\newcommand{\bbR}{\mathbb{R}}

\def \mcb {{\mathcal B}}
\def \mcc {{\mathcal C}}

\def \mci {{\mathcal I}}

\def \mcm {{\mathcal M}}

\def \mcr {{\mathcal R}}

\def \mcu {{\mathcal U}}
\def \mcv {{\mathcal V}}

\def \mbr {{\mathbb R}}
\def \mbs {{\mathbb S}}

\def \id {\operatorname{Id}}

\def \supp {\text{supp }}

\def \beqq {\begin{equation}}
\def \eeqq {\end{equation}}
\def \bpf {\begin{proof}}
\def \epf {\end{proof}}
\def \beq {\begin{equation*}}
\def \eeq {\end{equation*}}

\def \eps {\epsilon}

\def \lap {\Delta}
\def \p {\partial}
\def \ha {\frac{1}{2}}   

\begin{document}
\title[]{Reconstruction of the observable universe from the integrated Sachs-Wolfe effect}
\author{Julianne Chung and Yiran Wang}
\address{Julianne Chung
\newline
\indent Department of Mathematics, Emory University}
\email{jmchung@emory.edu}
\address{Yiran Wang
\newline
\indent Department of Mathematics, Emory University}
\email{yiran.wang@emory.edu}

\begin{abstract}
The integrated Sachs-Wolfe (ISW) effect is a property of the Cosmic Microwave Background (CMB), in which photons from the CMB are gravitationally redshifted, causing the anisotropies in the CMB. An intriguing question is whether one can infer the gravitational perturbations from the ISW effect observed near the Earth. In this work, we address the question using a tomographic reconstruction approach, similar to X-ray CT reconstruction in medical imaging. We develop the mathematical analysis for the stable inversion of the X-ray transform in the cosmological setting. 
In addition, we provide a numerical study of reconstruction methods, thereby demonstrating the feasibility and potential of the tomography method. 
\end{abstract}
\date{\today}
\maketitle
\tableofcontents

\section{The physical  background} 
This paper concerns the reconstruction of the initial status of the universe from the Cosmic Microwave Background (CMB). We start with the description of the physical problem. Consider the Friedman-Lema\^ite-Robertson-Walker (FLRW) model for the universe:
\beq
\mcm = (0, \infty)\times \mbr^3, \ \ g_0 = -dt^2 + a^2(t)dx^2,
\eeq
where $t \in (0, \infty), x\in \mbr^3$. The factor $a(t)$ is assumed to be positive and smooth in $t$. It represents the rate of expansion of the universe. We think of $(\mcm, g_0)$ as the background universe, which is isotropic and homogeneous. We note that the metric is conformal to the Minkowski metric. More precisely, after a change of variables from $(t, x)$ to $(s, x), s>0$ with $ds/dt = (a(t))^{-1}$,  we can write the metric as $g_0 = a^2(s)( -ds^2  + dx^2)$.

Next, we consider the actual universe $(\mcm, g)$ as a perturbation of $(\mcm, g_0)$. The CMB perturbation theory is well-developed in the literature, see for instance \cite{Dod, Dur, MFB}. Here, we consider a universe governed by a scalar field $\phi$. The stress energy tensor is  
\beq
T^\mu_{\ \ \nu} = \nabla^\mu \phi \nabla_\nu \phi - [\ha \sum_{\alpha = 0}^3 \nabla^\alpha\phi \nabla_\alpha\phi - V(\phi)]\delta^\mu_{\ \ \nu}, \quad \mu, \nu = 0, 1, 2, 3, 
\eeq
see \cite[equation (6.2)]{MFB}. Here, $V$ is the potential function for the scalar field $\phi$. The field itself satisfies the Klein-Gordon equation $\square \phi + \p_\phi V(\phi) = 0.$  According to Einstein's relativity theory, $g$ satisfies the Einstein equation with the source $T$. We work with the linearized theory and assume that $\phi = \phi_0 + \delta \phi$ where $\phi_0$ is the scalar field which drives the background model and $\delta \phi$ denotes the perturbation. Then we can split $T^\mu_{\ \ \nu} = {}^{(0)}T^\mu_{\ \ \nu} + \delta T^{\mu}_{\ \ \nu}$. From the linearized Einstein-Klein-Gordon equations, one deduces that the metric $g$ up to the  first order perturbation is of the form
\beqq\label{eq-confg}
g = a^2(s)[ (1+ 2\Phi) ds^2 - (1-2\Phi) dx^2],
\eeqq
where $\Phi$ is a scalar functions on $\mcm$. In fact, $\Phi = \Psi$  satisfies the equation 
\beqq\label{eq-eqscalar}
\Phi'' + 2(H - \phi_0''/\phi_0') \Phi' - \lap \Phi + 2(H' - H \phi_0''/\phi_0)\Phi = 0,
\eeqq
see \cite[equation (6.48)]{MFB}. Here, $H(s) = a'(s)/a(s)$ and $'$ denotes the derivative in $s$. Note that \eqref{eq-eqscalar} is a damped wave equation. We remark that for a universe dominated by perfect fluids, there is a similar model where $\Phi$ satisfies the Bardeen's equation which is a wave equation with sound speed $c\in (0, 1)$, see for example \cite{MFB, VaWa}. We consider the scalar field model because we are interested in finding gravitational waves which travel at the speed of light. The scalar field model can be regarded as a simplified problem, in addition to its own interest. 

Now we consider the photon distribution in such a universe. It is known that trajectories of photons can be represented by light-like (or null) geodesics for the Lorentzian metric $g$ on $\mcm$.  Let $\mcm_0 = \{s_0\}\times \mbr^3$ be the surface of last scattering.  This is the moment after which photons stopped interacting and started to travel freely in $\mcm.$  Let $\mcm_1 = \{s_1\}\times \mbr^3, s_1>s_0$ be the surface where we make observations of the photons or the CMB. We consider the photon energies observed at $\mcm_0, \mcm_1$,   
\beq
E_0 = g_0(\dot \gamma(\tau_0), \p_s)\quad  \mbox{and} \quad E_1 = g_0(\dot \gamma(\tau_1), \p_s).
\eeq
Here, $\gamma(\tau), \tau \in \mbr$ is a light-like geodesic with $\gamma(\tau_0)\in \mcm_0, \gamma(\tau_1)\in \mcm_1$. Also, the flow of the vector field $\p_s$ represents the observer. The redshift $z$ is defined by 
\beq
1 + z = E_1/E_0.
\eeq
In \cite{SaWo}, Sachs and Wolfe derived that to the first order linearization, the redshift is represented by an integral  of the metric perturbations, see \cite[equation (39)]{SaWo} and also the derivation in \cite{LOSU}. This is known as the integrated Sachs-Wolfe (ISW) effect.   In our setup of scalar perturbations, the ISW is given by (up to some scalar factors),
\beqq\label{eq-sw}
\begin{gathered}
\int  \p_s \Phi(\gamma(\tau))  d\tau 
\end{gathered}
\eeqq
 where $\gamma(\tau)$ is a light-like geodesic from $\mcm_0$ to $\mcm_1$. See also \cite[Section 2.5]{Dur} and \cite{Dod}.

 \begin{figure}[t]
\includegraphics[width =.95 \textwidth]{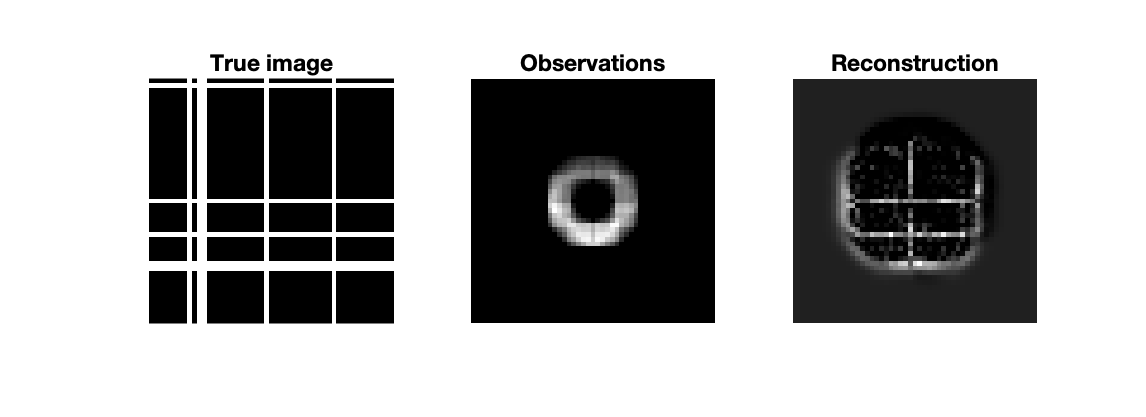}
\caption{
Cosmological X-ray tomography for a 2D Universe. The left figure is the true image which represents a collection of cosmic strings or gravitational plane waves at time $t=0$. The middle figure shows the simulated ISW effect observed at $t=T$ by several detectors near the center of the image. One can see the ``anisotropies" but the line structures are hardly discernible. The right figure shows the reconstruction using the tomography method in this paper. In the region given by Theorem \ref{thm-main}, the line structures are clearly visible. 
}
\label{fig-demo}
\end{figure}

It is clear from \eqref{eq-sw} that the ISW contains information of the gravitational perturbations. An outstanding question is whether one can extract the information. 
Since the discovery of the CMB by Penzias and Wilson in 1963, the measurements of the CMB have been significantly improved through multiple satellite projects such as COBE (Cosmic Background Explorer), WMAP (Wilkinson Microwave Anisotropy Probe), and the Planck Surveyor. Even though ISW cannot be directly read off from the CMB data, there are many studies on extracting ISW from the CMB by combining it with other survey data. See \cite{Sha, Dup, MaDo, MuHu} for example.  There is a huge literature on recovering cosmological parameters from the CMB and ISW, see \cite{Dod, Dur} and \cite{Kab, Kro} for example. Most of the existing work is based on the analysis of the CMB power spectrum and statistical inference methods. In this work, we treat the problem from a tomography point of view, which is very similar to X-ray computed tomography in medical imaging. One advantage of the tomography approach is the capability to reconstruct detailed object structures. To illustrate this point,  we showcase the tomography reconstruction for a 2D universe in Figure \ref{fig-demo} from one numerical simulation in Section \ref{sec-num}. 

The transform \eqref{eq-sw} (for tensors) appeared in \cite{Gui} for the mathematical study of a cosmology problem. On page 186 of \cite{Gui}, Guillemin provisionally called the problem ``cosmological X-ray tomography". Recently, the relevance of the transform and the CMB inverse problem started to attract attention, see \cite{Wan2} for a survey. In particular, using microlocal methods, the authors of \cite{LOSU} demonstrated that time-like singularities (in the sense of wave front set) of $\Phi$ can be recovered from the ISW. See also \cite{Wan} for related results on light-like singularities. Unfortunately, not all information can be stably recovered by using only the transform, which is an issue already pointed out in \cite{Gui}. We refer to \cite{COW} for a recent numerical demonstration of the issue. A critical step was made in \cite{VaWa} that by taking into account the physical model, in particular the evolution equation \eqref{eq-eqscalar}, $\Phi$ can be stably recovered with the observation of \eqref{eq-sw} on a Cauchy surface, see also \cite{Wan1}. The result suggests that the tomography approach is promising, but to make it applicable,  one must address the problem in a practical setting where the CMB is observed only {\em near} the Earth.  The goal and novelty of this work is to address this problem by solving a cosmological X-ray tomography problem with partial data. We also conduct numerical studies to demonstrate the feasibility of the tomography method for cosmological applications. 

\section{The mathematical formulation and main results}\label{sec-main}
\begin{figure}[t]
\centering
\includegraphics[scale = 0.5]{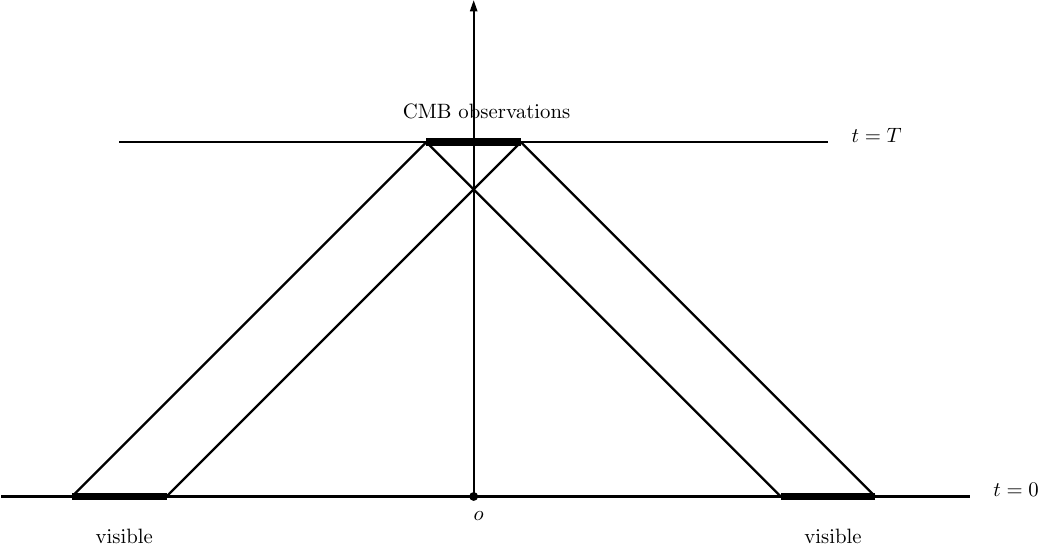}
\caption{The setup of the partial data problem. The marked set on $t = 0$ is the visible set in Theorem \ref{thm-main}.}
\label{fig-setup}
\end{figure}

For $T>0$, let $\mcm = [0, T]\times \mbr^3$ and $(t, x), t\in [0, T], x\in \mbr^3$ be the local coordinates. Let  $g = -dt^2 + \sum_{i = 1}^3dx_i^2$ be the Minkowski metric on $\mcm$.  We study the problem for a 3D universe for physical relevance, but remark that our method applies to dimensions $n\geq 2.$ For $t\in \mbr,$ we denote $\mcm_t = \{t\}\times \mbr^3$.  Let $\mcb  = \{x\in \mbr^3: |x| < 1\}$ be the unit ball on $\mbr^3$. We consider null geodesics on $(\mcm, g)$ from $\mcm_0$ that meet $\mcb$ on $\mcm_T$. See Figure \ref{fig-setup}. This setup corresponds to making CMB observations near the Earth. It is convenient to parametrize the null geodesics using $y\in \mcb$ and $v\in \mbs^2$ as  
\beqq\label{eq-geo}
\gamma_{y, v}(t) = (t, y + tv - Tv), \quad t\in \mbr.
\eeqq
For $t\in [0, T]$, $\gamma_{y, v}(t)$ is contained in $\mcm$. Note that the set of such null geodesics can be identified with the set $\mcc = \mcb \times \mbs^2.$ For a scalar function $f$ in $\mcm$, we consider the light ray transform  of $f$ defined as   
\beqq\label{eq-lray}
Lf(y, v) = \int_{0}^{T}f(t, y+ tv - Tv) dt, 
\eeqq
which is a function on $\mcc.$  

We study $L$ acting on solutions to the Cauchy problem of wave equations on $\mcm$. 
 Denote $\square = \p_t^2 +   \lap $ where $\lap$ is the positive Laplacian on $\mbr^3,$ namely $\lap = \sum_{i = 1}^3 D_{x_i}^2, D_{x_i} = -\imath\frac{\p}{\p x_i}, \imath = \sqrt{-1}.$  Let 
 \beqq\label{eq-P}
 P(t, x, \p_t, D) = \square + a_0(t, x) \p_t + \sum_{j = 1}^3 a_j(t, x)D_j + b(t, x), 
 \eeqq
 where $a_j, b\in C^\infty.$   We consider the Cauchy problem
\beqq\label{eq-cau}
\begin{gathered}
P(t, x, \p_t, D)u(t, x) = 0, \ \  t > 0, x\in \mbr^3 \\
u(0, x) = f_1(x), \ \ \p_t u(0, x) = f_2(x).
\end{gathered}
\eeqq
It is well-known that under proper regularity assumptions on $f_1, f_2,$ there is a unique solution $u$. We study the inverse problem of recovering information of $f_1, f_2$ from $L u$. 

For the full data problem namely $\mcb = \mbr^3$, it was shown in \cite{VaWa} (see also \cite{Wan1}) that one can stably recover $f_1, f_2$ supported in a fixed compact set. The partial data problem we are considering is different and it is not always possible to recover $f_1, f_2$.  Consider the setup in Figure \ref{fig-setup}. Let $\mcb_{1+T} = \{x\in \mbr^3: |x| <1+ T\}$ on $\mcm_0$. If $f_1, f_2$ are supported outside of $\mcb_{1+T}$, by finite speed of propagation of \eqref{eq-cau}, we see that $\supp u$ does not meet any light rays that intersect $\mcb$ at $t = T$. So $Lu = 0$. Thus the best one can hope for is to recover $f_1, f_2$ in $\mcb_{1+T}.$

For $b>a>0$, we set 
\beq
\mcr_{a, b} = \{x\in \mbr^3: a< |x|< b\}. 
\eeq
We use the convention that if $a\leq 0$ then $\mcr_{a, b} = \mcb_{b}$. In this work, we show that Cauchy data supported in $\mcr_{T-1, T+1}$ can be stably recovered from the observation on $\mcb.$ 
\begin{theorem}\label{thm-main}
Let $s\geq 0$ be an integer. Suppose $f_1 \in H^{s+1}(\mbr^3), f_2\in H^{s}(\mbr^3)$ are compactly supported in $\mcr_{T-1, T+1}$. Then $f_1, f_2$ are uniquely determined by $Lu$. Moreover, we have the following stability estimates 
\beqq\label{eq-mainest}
\|f_1\|_{H^{s+1}} + \|f_2\|_{H^s} \leq C \|Lu\|_{H^{s+2}} 
\eeqq
where $C>0$ is uniform for $f_1, f_2$ supported in a fixed compact set of $\mcr_{T-1, T+1}$. 
 \end{theorem}

We have two immediate remarks. First, in view of the stability estimate \eqref{eq-mainest}, the result can be generalized to small metric perturbations as in \cite{VaWa}. Second, if the Cauchy data is supported in $\mcb_{T+1}$ but not necessarily in $\mcr_{T-1, T+1}$, our analysis can potentially tell what singularities (wave front sets) of the Cauchy data can possibly be recovered, see Section \ref{sec-pf1} especially the microlocal inversion formula \eqref{eq-q10}, \eqref{eq-q11}. We do not pursue it further in this paper. 

The rest of the paper is organized as follows. In Section \ref{sec-back}, we study the microlocal properties of the back-projection. In Section \ref{sec-pf1}, we construct the microlocal inversion for the back-projection. Then we prove Theorem \ref{thm-main} in Section \ref{sec-pf}. Finally, we conduct a numerical study in Section \ref{sec-num}, and conclusions are provided in Section \ref{sec-conc}.

\section{Analysis of the backprojection}\label{sec-back} 
We start with the expression of the solution for the Cauchy problem \eqref{eq-cau}. It is well-known that one can find approximate solutions as oscillatory integrals, see for example \cite[Section 1, Chapter VI]{Tre}. Let $E_j(t), j = 1, 2$ be the fundamental solutions such that $P(t, x, \p_t, D)E_j(t) = 0$ for $t\in (0, T)$ and 
\beq
\begin{gathered} 
E_1(0) = \id, \quad \p_tE_1(0) = 0,   \\
E_2(0) = 0, \quad \p_t E_2(0) = \id,
\end{gathered}
\eeq
where $\id$ denotes the identity operator.  Then we can write the solution of \eqref{eq-cau} as $u(t, x) = E_1(t)f_1 + E_2(t)f_2$. In fact, $E_j(t), j = 1, 2$ are of the form  
\beqq\label{eq-fund}
\begin{gathered}
E_j(t) u(x) = E_j^+(t) u(x) + E_j^-(t)u(x)  + R_j(t)u(x)
\end{gathered}
\eeqq
where 
\beqq\label{eq-E}
\begin{gathered}
E_j^+(t)u(x) = (2\pi)^{-3} \int_{\mbr^3}\int_{\mbr^3} e^{\imath ((x-y)\cdot \xi + t |\xi|)} a_{j}^+(t, x, \xi) u(y) dy d\xi\\
E_j^-(t)u(x) =  (2\pi)^{-3} \int_{\mbr^3}\int_{\mbr^3} e^{\imath ((x-y)\cdot \xi - t |\xi|)} a_{j}^-(t, x, \xi)u(y) dy d\xi 
\end{gathered}
\eeqq
and $R_j(t), j = 1, 2$ are regularizing, see \cite[Chapter VI, (1.37)]{Tre}. Here, the phase function is derived from the principal part of $P(t, x, \p_t, D)$. The amplitudes $a_j^\pm$ have asymptotic summations of the form   
\beq
a_j^\pm(t, x, \xi) \sim \sum_{k = 0}^\infty a_{jk}^\pm(t, x, \xi),
\eeq
where each $a_{jk}^\pm$ is homogeneous of degree $-j-k$. They are determined via transport equations. Here, we recall the construction of the leading order terms for $k = 0$. They satisfy
\beqq\label{eq-trans}
\p_t a_{j0}^\pm \mp \sum_{l = 1}^3 \frac{\xi_l}{|\xi|}  \p_l a_{j0}^\pm \pm \frac{1}{2 } (a_0(t, x) + \sum_{l = 1}^3 a_l(t, x)\frac{\xi_l}{|\xi|}) a_{j0}^\pm = 0, 
\eeqq
see \cite[Chapter VI, (1.49)]{Tre}. The initial condition at $t = 0$ is given by (see \cite[Chapter VI, (1.53)]{Tre})
\beq
\begin{gathered}
(\imath |\xi|)^{j'-1} a_{j0}^+(0, x, \xi) +(- \imath |\xi|)^{j'-1} a_{j0}^-(0, x, \xi) = \delta_{jj'}, 
\end{gathered}
\eeq
where $j, j' = 1, 2$ and $\delta_{jj'}$  is the Kronecker delta function.  We find that 
\beq
a_{10}^\pm(0, x, \xi) = \ha, \quad a_{20}^\pm(0, x, \xi) = \pm\frac{1}{2\imath} |\xi|^{-1}.
\eeq
Then we can solve \eqref{eq-trans} with these initial conditions to get $a_{j0}^{^\pm}(t, x, \xi)$. In particular, we see that $a_{j0}^\pm$ are non-vanishing. 

To summarize, for $t>0$, $E^\pm_j(t)$ $\in I^{- j+1}(\mbr^3\times\mbr^3; (C^\pm)')$ are elliptic Fourier integral operators in H\"ormander's notation (see Definition 25.2.1 of \cite{Ho4}), where the canonical relations are 
\beqq\label{eq-cano0}
\begin{gathered}
C^\pm_t = \{(x, \eta; y, \xi) \in T^*\mbr^3 \backslash 0\times T^*\mbr^3\backslash 0:  x = y -  t (\pm \xi/|\xi|), \eta  = \xi\}. 
\end{gathered}
\eeqq
Finally, we write the approximate solution of \eqref{eq-cau}  as 
\beqq\label{eq-u}
u(t, x) = E_1^+(t)f_1 + E_1^-(t)f_1 + E_2^+(t)f_2 + E_2^-(t)f_2 + R_1(t)f_1 + R_2(t)f_2
\eeqq
where $t\geq 0$. \\

We analyze $Lu$ with $u$ in \eqref{eq-u}. It suffices to consider the operator $(L E_j^\pm), j = 1, 2$ defined by $(LE_j^\pm)f(x, v) =  L (E_j^{\pm}(t) f)(x, v)$ for a scalar function $f$ on $\mbr^3.$ In order to ``invert" the operators to get $f$, we use the idea of backprojection in X-ray tomography. In particular, we follow the approach of \cite{VaWa} to take the integral of the $v$ variable on $\mbs^2$.  For $h\in C^\infty(\mbr^3\times \mbs^2),$ we define 
\beq
I  h(x) = \int_{\mbs^2}  h(x, v) dv.
\eeq 
The main purpose of this section is to analyze the microlocal structure of $I L E^\pm_j$ and show that it can be decomposed into a sum of FIOs. The main results of this section are Lemma \ref{lm-b12} and \ref{lm-b34}. 

\begin{figure}[t]
\centering
\includegraphics[scale = 0.5]{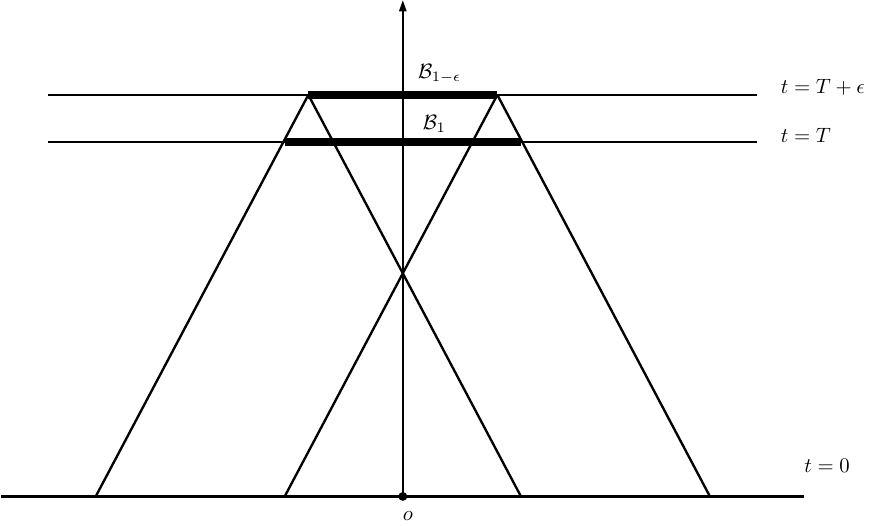}
\caption{Illustration of the modified light ray transform $L_\eps$. }
\label{fig-pf}
\end{figure}

To avoid some technicalities, we will use a subset of the light rays. For $\eps>0$ small, we consider the set $\mcb_{1-\eps} = \{y\in \mbr^3: |y|< 1-\eps\}$. Then we use the light rays that meet $\{T+\eps\}\times \mcb_{1-\eps}$, see Figure \ref{fig-pf}. We let $\phi_\eps$ be a smooth cut-off function on $\mbr^3$ such that $\supp \phi_\eps =  \overline \mcb_{1-\eps}$, $\phi_\eps>0$ on $\mcb_{1-\eps}$ and $\phi_\eps = 1$ on $\overline\mcb_{1-2\eps}$. Then we let $L_\eps = \phi_\eps L$ so that 
\beqq\label{eq-lraye}
L_\eps f(y, v) = \int_0^T \phi_\eps(y) f(t, y + tv - (T +\eps)v)dt, \quad y\in \mcb_{1}, v\in \mbs^2. 
\eeqq
Note that when $t = T$, $y+tv-(T+\eps)v = y- \eps v \in \mcb$ for $y\in \mcb_{1-\eps}.$ The role of   $\phi_\eps$ is to avoid the cut-off singularities at the boundary of $\mcb_1$. 

We now analyze $I  L_\eps E_1^+ $. In particular, we will find the integral kernel of the operator in the sense of distributions. Using \eqref{eq-E}, we have 
\beq
\begin{gathered}
 I L_\eps E_1^+ f(x) 
 = (2\pi)^{-3} \int_{\mbs^2} \int_{0}^T\int_{\mbr^3}\int_{\mbr^3}  e^{\imath(x - y)\cdot \xi} e^{\imath (t - T -\eps)v\cdot \xi} e^{\imath t |\xi|}a_1^+(t, x, \xi)   \phi_\eps(x) f(y) dy d\xi dt dv.  
  \end{gathered}
  \eeq
Note that for $t\in (0, T)$, $t - T - \eps < 0$. For $|\xi|$ large, the phase function
\beq
(t-T-\eps)v\cdot \xi = (t-T-\eps)|\xi| v\cdot(\xi/|\xi|)
\eeq
has non-degenerate critical points at $v = \pm \xi/|\xi|$, see e.g., \cite[(1.14)]{Mel}. We can perform the stationary phase method for the integration in $v$ to get $IL_\eps E_1^+ = B_{1, 1}^+ + B_{1, 2}^+ + R_1^+$ where $R_1^+$ is regularizing and   
  \beqq\label{eq-B1}
\begin{gathered}
 B_{1, 1}^+ f(x) =  (2\pi)^{-3} \int_{0}^T\int_{\mbr^3 } \int_{\mbr^3 }e^{\imath(x - y)\cdot \xi} e^{\imath (T+\eps)|\xi|}  b_{1, 1}^+(t, x, \xi)  \phi_\eps(x)  f(y) dy d\xi dt, \\
B_{1, 2}^+ f (x) = (2\pi)^{-3}  \int_{0}^T\int_{\mbr^3 } \int_{\mbr^3 }e^{\imath(x - y)\cdot \xi} e^{-\imath (T+\eps)|\xi|} e^{\imath 2 |\xi|t} b_{1, 2}^+(t, x, \xi)  \phi_\eps(x)   f(y) dy d\xi dt. 
\end{gathered}
\eeqq 
Here, $b_{1, k}^+, k = 1, 2$ are smooth functions supported away from $\xi = 0$. They have asymptotic summations of the form 
\beq
b_{1, k}^+(t, x, \xi) \sim \sum_{l = 1}^\infty |\xi|^{-l} b_{1, k; l}^+(t, x, \xi/|\xi|) 
\eeq
for $|\xi|$ large. The leading order terms are 
\beqq\label{eq-symb}
\begin{gathered}
 \sigma_{-1}(b_{1, 1}^+)(t, x, \xi) 
= C_1 \frac{1}{(t - T - \eps)|\xi|} a_{10}^+(t, x, \xi/|\xi|) =  \frac{C_1}{2(t - T - \eps)|\xi|}, \\
 \sigma_{-1}(b_{1, 2}^+)(t, x, \xi) 
= C_2 \frac{1}{(t - T - \eps)|\xi|} a_{10}^{+}(t, x, -\xi/|\xi|) = \frac{C_2}{2(t - T - \eps)|\xi|}, 
\end{gathered}
\eeqq
where we used $\sigma_{-1}(\cdot)$ to denote the leading order term or the principal symbol of order $-1$, and $C_1, C_2$ come from the constants in the stationary phase argument, see  \cite[(1.14)]{Mel}. More explicitly,  
\beq
\begin{gathered}
C_1 =  e^{-\pi \imath/2} (2\pi)^2   =  -4\pi^2 \imath, \quad 
 C_2 =  e^{\pi \imath/2} (2\pi)^2   = 4\pi^2 \imath. 
 \end{gathered}
\eeq

Now we can write $B_{1,1}^+$ as 
\beqq\label{eq-B11}
\begin{gathered}
B_{1, 1}^+ f(x)    = (2\pi)^{-3}  \int_{\mbr^3}\int_{\mbr^3 } e^{\imath(x - y)\cdot \xi} e^{\imath (T+\eps)|\xi|} \tilde b_{1, 1}^+(x, \xi)  f(y) dy d\xi  
\end{gathered}
\eeqq 
where 
\beq
\tilde b_{1, 1}^+(x, \xi) = \int_0^T  b_{1, 1}^+(t, x, \xi) \phi_\eps(x) dt
\eeq
is a symbol of order $-1$. The leading order term is given by
\beqq\label{eq-bplus-sym}
\sigma_{-1}(\tilde b_{1, 1}^+)(x, \xi) = |\xi|^{-1}\phi_\eps(x) \int_0^T  \frac{-4\pi^2\imath}{2(t - T - \eps)}  dt = (-2\pi^2\imath \ln\frac{\eps}{T+\eps}) |\xi|^{-1} \phi_\eps(x) 
\eeqq 
for $|\xi|$ large. 
We observe that  $B_{1, 1}^+$ is an Fourier integral operator of order $-1$ with canonical relation $C_{T+\eps}$. Hereafter, for $a\in \mbr$, we define 
\beq
\begin{gathered}
C_a = \{(x, \xi; y, \eta): \xi = \eta,  x= y - a \xi/|\xi|,  
x, y \in \mbr^3, \xi, \eta \in \mbr^3\backslash 0 \}. 
\end{gathered}
\eeq
In H\"ormander's notation, we can write $B_{1, 1}^+\in I^{-1}(\mbr^3\times \mbr^3; C_{T+\eps}')$.

For $B_{1, 2}^+$ in \eqref{eq-B1}, we first compute  
\beq
\begin{gathered}
 \tilde b_{1, 2}^+(x, \xi) = \int_0^T e^{\imath 2 |\xi|t} b_{1, 2}^+(t, x, \xi)\phi_\eps(x) dt 
  =   \frac{1}{\imath 2 |\xi|}  b_{1, 2}^+(T, x, \xi)\phi_\eps(x) e^{\imath 2 |\xi|T} \\ -  \frac{1}{\imath 2 |\xi|}  b_{1, 2}^+(0, x, \xi)\phi_\eps(x)   -  \int_0^T  \frac{1}{\imath 2 |\xi|}  \p_t b_{1, 2}^+(t, x, \xi)\phi_\eps(x)  e^{\imath 2 |\xi|t} dt
\end{gathered}
\eeq
via integration by parts. This procedure can be continued to get 
\beq
 \tilde b_{1, 2}^+(x, \xi)  
  = \tilde b^+_{1, 2, 1}(x, \xi) + e^{\imath 2 |\xi|T} \tilde b_{1, 2, 2}^+(x, \xi)  
  \eeq
where $\tilde b^+_{1, 2, j}, j = 1, 2$ are symbols of order $-2$. In fact, we can find the leading order terms as
\beq
\begin{gathered}
\sigma_{-2}(\tilde b^+_{1, 2, 1})(x, \xi) = \frac{\pi^2}{(T+\eps)}|\xi|^{-2}\phi_\eps(x), \quad 
\sigma_{-2}(\tilde b^+_{1, 2, 2})(x, \xi) = -\frac{\pi^2}{\eps}|\xi|^{-2}\phi_\eps(x). 
\end{gathered}
\eeq
Therefore, we have 
\beq
\begin{gathered}
B_{1, 2}^+  f(x)   = (2\pi)^{-3}  \int_{\mbr^3}\int_{\mbr^3 } e^{\imath(x - y)\cdot \xi} e^{-\imath (T+\eps)|\xi|}   \tilde b_{1, 2, 1}^+(x, \xi)  f(y) dy d\xi\\ 
    + (2\pi)^{-3}  \int_{\mbr^3}\int_{\mbr^3 } e^{\imath(x - y)\cdot \xi} e^{\imath (T-\eps)|\xi|}   \tilde b_{1, 2, 2}^+(x, \xi)   f(y) dy d\xi 
     = B_{1, 2, 1}^+f(x) + B_{1, 2, 2}^+f(x). 
\end{gathered}
\eeq 
We observe that the operators $B_{1, 2, k}^+, k = 1, 2$ are FIOs of order $-2$ with canonical relations  $C_{-T-\eps}$ and $C_{T-\eps}$. This completes the analysis for $IL_\eps E_1^+$. In particular, we decomposed the operator into a sum of FIOs. 

The analysis for $IL_\eps E_2^+$ is completely identical. The only difference is  the order and the symbol. We will not show the details. To summarize, we proved 
\begin{lemma}\label{lm-b12}
For $l = 1, 2$, we can write 
\beq
I L_\eps E_l^+ = B_{l, 1}^+ + B_{l, 2, 1}^+ + B_{l, 2, 2}^+ + R_l^+, 
\eeq
where $R_l^+$ are regularizing operators and the rest are Fourier integral operators 
\beq
\begin{gathered}
B_{l, 1}^+  \in I^{-1 - (l-1)}(\mbr^3\times \mbr^3; C'_{T+\eps}), \\
B_{l, 2, 1}^+ \in  I^{-2 - (l-1)}(\mbr^3\times \mbr^3; C'_{-T - \eps}),  \quad B_{l, 2, 2}^+ \in I^{-2 - (l-1)}(\mbr^3\times \mbr^3; C'_{T - \eps}). 
\end{gathered}
\eeq
Moreover,  
\begin{enumerate}
\item $B_{l, 1}^+, l = 1, 2$ can be written as an oscillatory integrals 
 \beq 
\begin{gathered}
B_{l, 1}^+ f(x)   = (2\pi)^{-3}  \int_{\mbr^3 }   \int_{\mbr^3 } e^{\imath(x - y)\cdot \xi} e^{\imath (T+\eps)|\xi|} \tilde b_{l, 1}^+(x, \xi) f(y)dy d\xi, 
\end{gathered}
\eeq
where $\tilde b_{l, 1}^+$ are symbols of order $-1-(l-1)$. The principal symbols of $B_{l, 1}^+$ in this representation  are given by 
\beq
\begin{gathered}
\sigma_{-1}(B_{1, 1}^+)(x, y, \xi) = (-2\pi^2\imath \ln\frac{\eps}{T+\eps})  \phi_\eps(x) |\xi|^{-1},  \\
 \sigma_{-2}(B_{2, 1}^+)(x, y, \xi) = (-2\pi^2  \ln\frac{\eps}{T+\eps})  \phi_\eps(x)|\xi|^{-2}
\end{gathered}
\eeq
for $|\xi|$ large. 

\item  $B_{l, 2, 1}^+, l = 1, 2$ can be written as an oscillatory integrals 
 \beq 
\begin{gathered}
B_{l, 2, 1}^+ f(x)   = (2\pi)^{-3}  \int_{\mbr^3 }   \int_{\mbr^3 } e^{\imath(x - y)\cdot \xi} e^{-\imath (T+\eps)|\xi|} \tilde b_{l, 2, 1}^+(x, \xi) f(y)dy d\xi, 
\end{gathered}
\eeq
where $\tilde b_{l, 2, 1}^+$ are symbols of order $-2-(l-1)$. The principal symbols of $B_{l, 2, 1}^+$ in this representation  are given by 
\beq
\begin{gathered}
\sigma_{-2}(B_{1, 2, 1}^+)(x, y, \xi) =  \frac{\pi^2}{(T+\eps)}\phi_\eps(x)|\xi|^{-2},   \quad 
 \sigma_{-3}(B_{2, 2, 1}^+)(x, y, \xi) =  \frac{\pi^2}{\imath (T+\eps)} \phi_\eps(x) |\xi|^{-3}
\end{gathered}
\eeq  
for $|\xi|$ large.

\item $B_{l, 2, 2}^+, l = 1, 2$ can be written as an oscillatory integrals 
 \beq 
\begin{gathered}
B_{l, 2, 2}^+ f(x)   = (2\pi)^{-3}  \int_{\mbr^3 }   \int_{\mbr^3 } e^{\imath(x - y)\cdot \xi} e^{\imath (T-\eps)|\xi|} \tilde b_{l, 2, 1}^+(x, \xi) f(y)dy d\xi, 
\end{gathered}
\eeq
where $\tilde b_{l, 2, 2}^+$ are symbols of order $-2-(l-1)$. The  principal symbols of $B_{l, 2, 2}^+$ in this representation are given by 
\beq
\begin{gathered}
\sigma_{-2}(B_{1, 2, 2}^+)(x, y, \xi) =  -\frac{\pi^2}{\eps} \phi_\eps(x)  |\xi|^{-2},  \quad  
 \sigma_{-3}(B_{2, 2, 2}^+)(x, y, \xi) =  -\frac{\pi^2}{\eps\imath } \phi_\eps(x) |\xi|^{-3}
\end{gathered}
\eeq
for $|\xi|$ large. 

\end{enumerate}
\end{lemma}

We can obtain a similar decomposition for $I L_\eps E_j^- $ following the same type of calculation. Because the canonical relations are  different, we repeat some of the calculations. Consider $IL_\eps E_1^-$. We compute that   
\beq
\begin{gathered}
I L_\eps E_1^- f(x)  
 = (2\pi)^{-3} \int_{\mbs^2} \int_{0}^T\int_{\mbr^3}\int_{\mbr^3}  e^{\imath(x - y)\cdot \xi} e^{\imath (t - T -\eps)v\cdot \xi}  e^{-\imath t |\xi|}a_1^-(t, x, \xi)   \phi_\eps(x) f(y) dy d\xi dt dv. 
  \end{gathered}
  \eeq
Again, for $|\xi|$ large, we perform the stationary phase method for the integration in $v$ to get $IL_\eps E_1^- = B_{1, 1}^- + B_{1, 2}^-+ R_1^-$ where $R_1^-$ is regularizing and   
  \beqq\label{eq-B2}
\begin{gathered}
B_{1, 1}^- f (x) =   (2\pi)^{-3} \int_{0}^T\int_{\mbr^3 } \int_{\mbr^3 }e^{\imath(x - y)\cdot \xi} e^{-\imath(T+\eps)|\xi|}  b_{1, 1}^-(t, x, \xi)  \phi_\eps (x)   f(y) dy d\xi dt, \\
 B_{1, 2}^- f(x) =   (2\pi)^{-3}  \int_{0}^T\int_{\mbr^3 } \int_{\mbr^3 }e^{\imath(x - y)\cdot \xi} e^{\imath (T+\eps)|\xi|} e^{-\imath 2|\xi|t}   b_{1, 2}^-(t, x, \xi)  \phi_\eps(x)  f(y) dy d\xi dt. 
\end{gathered}
\eeqq 
Similarly, $b_{1, k}^-, k = 1, 2$ are smooth functions supported away from $\xi = 0$. They have asymptotic summations of the form 
\beq
b_{1, k}^-(t, x, \xi) \sim \sum_{l = 1}^\infty |\xi|^{-l} b_{1, k, l}^-(t, x, \xi/|\xi|) 
\eeq
for $|\xi|$ large. The leading order terms are 
\beqq\label{eq-symb}
\begin{gathered}
\sigma_{-1}(b_{1, 1})(t, x, \xi) 
= C_1 \frac{1}{(t - T - \eps)|\xi|}  a_{10}^-(t, x, \xi/|\xi|) = \frac{C_1}{2(t - T - \eps)|\xi|}, \\
\sigma_{-1}(b_{1, 2})(t, x, \xi) 
= C_2 \frac{1}{(t - T - \eps)|\xi|}  a_{10}^{-}(t, x, -\xi/|\xi|) =\frac{C_2}{2(t - T - \eps)|\xi|}. 
\end{gathered}
\eeqq
We integrate in $t$ in $B_{1,1}^-$ to get 
\beq
\begin{gathered}
B_{1, 1}^- f(x)    =  (2\pi)^{-3}  \int_{\mbr^3}\int_{\mbr^3 } e^{\imath(x - y)\cdot \xi} e^{-\imath (T+\eps)|\xi|} \tilde b_{1, 1}^-(x, \xi)  f(y) dy d\xi, 
\end{gathered}
\eeq 
where 
\beq
\tilde b_{1, 1}^-(x, \xi) = \int_0^T |\xi|^{-1} b_{1, 1}^-(t, x, \xi) \phi_\eps(x) dt
\eeq
is a symbol of order $-1$. For $B_{1, 2}^-$, we obtain via integration by parts that
\beq
\begin{gathered}
 \tilde b_{1, 2}^-(x, \xi) = \int_0^T e^{-\imath 2 |\xi|t}|\xi|^{-1}  b_{1, 2}^-(t, x, \xi)\phi_\eps(x) dt\\
  =  \frac{1}{-\imath 2 |\xi|^2}  b_{1, 2}^-(T, x, \xi)\phi_\eps(x)  e^{-\imath 2 |\xi|T} -  \frac{1}{-\imath 2 |\xi|^2}  b_{1, 2}^-(0, x, \xi)\phi_\eps(x)  \\
  -  \int_0^T  \frac{1}{-\imath 2 |\xi|^2}  \p_t b_{1, 2}^-(t, x, \xi)\phi_\eps(x)  e^{-\imath 2 |\xi|t} dt. 
\end{gathered}
\eeq
The procedure can be continued  to yield
\beq
 \tilde b_{1, 2}^-(x, \xi)  
  = \tilde b^-_{1, 2, 1}(x, \xi) + e^{-\imath 2 |\xi|T} \tilde b_{1, 2, 2}^-(x, \xi), 
  \eeq
where $\tilde b^-_{1, 2, j}, j = 1, 2$ are symbols of order $-2$.   Therefore, we have 
\beq
\begin{gathered}
B_{1, 2}^-  f(x)   =  (2\pi)^{-3} \int_{\mbr^3}\int_{\mbr^3 } e^{\imath(x - y)\cdot \xi} e^{-\imath (T-\eps)|\xi|}   \tilde b_{1, 2, 1}^-(x, \xi)  f(y) dy d\xi\\ 
    +  (2\pi)^{-3} \int_{\mbr^3}\int_{\mbr^3 } e^{\imath(x - y)\cdot \xi} e^{\imath (T+\eps)|\xi|}   \tilde b_{1, 2, 2}^-(x, \xi)   f(y) dy d\xi 
     = B_{1, 2, 1}^-f(x) + B_{1, 2, 2}^- f(x). 
\end{gathered}
\eeq 
To summarize, we proved 
\begin{lemma}\label{lm-b34}
For $l = 1, 2$, we can write 
\beq
IL_\eps E_l^- = B_{l, 1}^- + B_{l, 2, 1}^- + B_{l, 2, 2}^- + R_l^-, 
\eeq
where $R_l^-$ are regularizing operators and the rest are elliptic Fourier integral operators 
\beq
\begin{gathered}
B_{l, 1}^-  \in I^{-1 -(l-1)}(\mbr^3\times \mbr^3; C'_{-T-\eps}), \\
B_{l, 2, 1}^-  \in I^{-2-(l-1)}(\mbr^3\times \mbr^3; C'_{-T + \eps}), \quad B_{l, 2, 2}^- \in I^{-2-(l-1)}(\mbr^3\times \mbr^3; C'_{T + \eps}). 
\end{gathered}
\eeq
Moreover,  
\begin{enumerate}
\item $B_{l, 1}^-, l = 1, 2$ can be written as oscillatory integrals, 
 \beq 
\begin{gathered}
B_{l, 1}^- f(x)   =  (2\pi)^{-3} \int_{\mbr^3 }   \int_{\mbr^3 } e^{\imath(x - y)\cdot \xi} e^{-\imath (T+\eps)|\xi|} \tilde b_{l, 1}^-(x, \xi) f(y)dy d\xi, 
\end{gathered}
\eeq
where $\tilde b_{l, 1}^-$ are symbols of order $-1-(l-1)$. The principal symbols of $B_{l, 1}^-$ in this representation are given by 
\beq
\begin{gathered}
\sigma_{-1}(B_{1, 1}^-)(x, y, \xi) = (-2\pi^2\imath \ln\frac{\eps}{T+\eps})  |\xi|^{-1} \phi_\eps(x), \\
 \sigma_{-2}(B_{2, 1}^-)(x, y, \xi) = (2\pi^2  \ln\frac{\eps}{T+\eps}) |\xi|^{-2} \phi_\eps(x), 
\end{gathered}
\eeq
for $|\xi|$ large. 
\item  $B_{l, 2, 1}^-, l = 1, 2$ can be written as oscillatory integrals, 
 \beq 
\begin{gathered}
B_{l, 2, 1}^- f(x)   =  (2\pi)^{-3} \int_{\mbr^3 }   \int_{\mbr^3 } e^{\imath(x - y)\cdot \xi} e^{\imath (-T+\eps)|\xi|} \tilde b_{l, 2, 1}^-(x, \xi) f(y)dy d\xi, 
\end{gathered}
\eeq
where $\tilde b_{l, 2, 1}^-$ are symbols of order $-2-(l-1)$. The principal symbols of $B_{l, 2, 1}^-$ in this representation  are given by 
\beq
\begin{gathered}
\sigma_{-2}(B_{1, 2, 1}^-)(x, y, \xi) =  -\frac{\pi^2}{(T+\eps)} \phi_\eps(x) |\xi|^{-2},  \quad 
 \sigma_{-3}(B_{2, 2, 1}^-)(x, y, \xi) =   -\frac{\pi^2\imath }{(T+\eps)} \phi_\eps(x)|\xi|^{-3}, 
\end{gathered}
\eeq  
for $|\xi|$ large. 

\item $B_{l, 2, 2}^-, l = 1, 2$ can be written as an oscillatory integrals 
 \beq 
\begin{gathered}
B_{l, 2, 2}^- f(x)   = (2\pi)^{-3}  \int_{\mbr^3 }   \int_{\mbr^3 } e^{\imath(x - y)\cdot \xi} e^{\imath (T+\eps)|\xi|} \tilde b_{l, 2, 1}^-(x, \xi) f(y)dy d\xi,   
\end{gathered}
\eeq
where $\tilde b_{l, 2, 2}^-$ are symbols of order $-2-(l-1)$. The  principal symbols of $B_{l, 2, 2}^-$ in this representation are given by 
\beq
\begin{gathered}
\sigma_{-2}(B_{1, 2, 2}^-)(x, y, \xi) =   \frac{\pi^2}{\eps}  \phi_\eps(x)  |\xi|^{-2},  \quad 
 \sigma_{-3}(B_{2, 2, 2}^-)(x, y, \xi) =   \frac{\pi^2\imath}{\eps} \phi_\eps(x) |\xi|^{-3}, 
\end{gathered}
\eeq
for $|\xi|$ large. 
\end{enumerate}
\end{lemma}

\section{The microlocal inversion} \label{sec-pf1}
For $\eps>0$ small, we consider $I L_\eps u$ with $u$ in \eqref{eq-u}. We apply Lemma \ref{lm-b12} and \ref{lm-b34} and group the terms as follows,
\beqq\label{eq-IL}
\begin{gathered}
  I L_\eps u  
=  \tilde B_{1}^+ f_1 + \tilde B_{2}^+ f_2 
+  B_{1, 2, 2}^+ f_1 + B_{2, 2, 2}^+ f_2\\
+  \tilde B_{1}^- f_1 + \tilde B_{2}^- f_2
+  B_{1, 2, 1}^- f_1 + B_{2, 2, 1}^- f_2 
+ R_1 f_1 + R_2 f_2, 
\end{gathered}
\eeqq
where $R_1 = R_1^+ + R_1^-, R_2 = R_2^+ + R_2^-$ are regularizing operators and 
\beqq\label{eq-oplist0}
\begin{gathered}
\tilde B_1^+ =  B_{1, 1}^+  + B_{1, 2, 2}^- \in I^{-1}(\mbr^3\times \mbr^3; C'_{T+\eps}),\\
\tilde B_2^+  =   B_{2, 1}^+ + B_{2, 2, 2}^- \in I^{-2}(\mbr^3\times \mbr^3; C'_{T+\eps}),  \\
\tilde B_1^- =  B_{1, 1}^-   + B_{1, 2, 1}^+ \in I^{-1}(\mbr^3\times \mbr^3; C'_{-T-\eps}), \\
  \tilde B_2^-  =   B_{2, 1}^- + B_{2, 2, 1}^+ \in I^{-2}(\mbr^3\times \mbr^3; C'_{-T-\eps}). 
\end{gathered}
\eeqq
Also, $B_{1, 2, 2}^+ \in I^{-2}(\mbr^3\times \mbr^3; C'_{T-\eps}), B_{2, 2, 2}^+\in I^{-3}(\mbr^3\times \mbr^3; C'_{T-\eps})$ and $B_{1, 2, 1}^- \in I^{-2}(\mbr^3\times \mbr^3; C'_{-T+\eps}), B_{2, 2, 1}^-  \in I^{-3}(\mbr^3\times \mbr^3; C'_{-T+\eps})$. Note that the orders of these FIOs acting on $f_1, f_2$ are lower than those corresponding terms in \eqref{eq-oplist0}, even though the canonical relations are different.  
The operators appearing in \eqref{eq-IL} are FIOs of graph type. Hence in principle, we can use FIO calculus (see for instance \cite[Section 25.3]{Ho4}) to solve for $f_1, f_2.$ However, equation \eqref{eq-IL} alone is not enough to solve for both $f_1$ and $f_2$. We will obtain another equation by changing the back-projection operator $I$, similar to the approach in \cite{VaWa}. 

Let $\varphi$ be a function on $\mbs^2$. Then for $h\in C^\infty(\mbr^3\times \mbs^2),$ we define 
\beq
I^\varphi  h(x) = \int_{\mbs^2} \varphi(v) h(x, v) dv.
\eeq 
We consider the back-projection $I^\varphi L_\eps u$. Ideally, we would like to take an odd function $\varphi$ such that $\varphi(-v) = -\varphi(v), v\in \mbs^2$ but then $\varphi$ would vanish somewhere on $\mbs^2$ so we proceed as follows. Let $x = (x_1, x_2, x_3)$ be the coordinate for $\mbr^3$.  For $\delta >0$, let $\mcu_k = \{v: v = (x_1, x_2, x_3), \|x\| = 1, |x_k| > \delta/2\}, k = 1, 2, 3.$ For $\delta$ sufficiently small, $\mcu_k, k = 1, 2, 3$ form an open covering of $\mbs^2.$ Let $\chi_k(v), k = 1, 2, 3$ be a partition of unity subordinated to this covering and $\chi_k(v) = 1$ on $\mcv_k = \{v: v = (x_1, x_2, x_3), \|x\| = 1, |x_k| > \delta\}, k = 1, 2, 3.$ Here, by possibly taking $\delta$ smaller, we can assume that $\mcv_k$ also forms an open covering of $\mbs^2.$ 
For $v\in \mbs^2$, we let 
\beq
\varphi_{k}(v) =  \chi_k(x) x_k + 2, \quad k = 1, 2, 3. 
\eeq
Then $\varphi_{k}(v)\neq 0$ and $\varphi_{k}(-v) - \varphi_{k}(v) \neq 0$ for $v\in \mcu_k.$ We now analyze $I^{\varphi_k} L_\eps u$ as in Section \ref{sec-back}. By repeating the calculations in Lemma \ref{lm-b12} and \ref{lm-b34}, we have the following Lemmas
\ref{lm-b12new} and \ref{lm-b34new}. The proofs are omitted. 
\begin{lemma}\label{lm-b12new}
For $k = 1, 2, 3,$ and $l = 1, 2$, we can write 
\beq
I^{\varphi_k} L_\eps E_l^+ = B_{l, 1}^{\varphi_k, +} + B_{l, 2, 1}^{\varphi_k, +} + B_{l, 2, 2}^{\varphi_k, +} + R_l^{\varphi_k, +}, 
\eeq
where $R_l^{\varphi_k, +}$ are regularizing operators and the rest are Fourier integral operators 
\beq
\begin{gathered}
B_{l, 1}^{\varphi_k, +}  \in I^{-1 - (l-1)}(\mbr^3\times \mbr^3; C'_{T+\eps}), \\
B_{l, 2, 1}^{\varphi_k, +} \in  I^{-2 - (l-1)}(\mbr^3\times \mbr^3; C'_{-T - \eps}),  \quad B_{l, 2, 2}^{\varphi_k, +} \in I^{-2 - (l-1)}(\mbr^3\times \mbr^3; C'_{T - \eps}). 
\end{gathered}
\eeq
Moreover,  
\begin{enumerate}
\item $B_{l, 1}^{\varphi_k, +}, l = 1, 2$ can be written as oscillatory integrals,
 \beq 
\begin{gathered}
B_{l, 1}^{\varphi_k, +} f(x)   = (2\pi)^{-3} \int_{\mbr^3 }   \int_{\mbr^3 } e^{\imath(x - y)\cdot \xi} e^{\imath (T+\eps)|\xi|} \tilde b_{l, 1}^{\varphi_k, +}(x, \xi) f(y)dy d\xi, 
\end{gathered}
\eeq
where $\tilde b_{l, 1}^{\varphi_k, +}$ are symbols of order $-1-(l-1)$. The principal symbols of $B_{l, 1}^{\varphi_k, +}$ in this representation  are given by 
\beq
\begin{gathered}
\sigma_{-1}(B_{1, 1}^{\varphi_k, +})(x, \xi) = (-2\pi^2\imath \ln\frac{\eps}{T+\eps})  \phi_\eps(x) \varphi_k(-\xi/|\xi|) |\xi|^{-1}, \\
 \sigma_{-2}(B_{2, 1}^{\varphi_k, +})(x, \xi) = (-2\pi^2  \ln\frac{\eps}{T+\eps})   \phi_\eps(x)\varphi_k(-\xi/|\xi|)|\xi|^{-2}, 
\end{gathered}
\eeq
for $|\xi|$ large. 
\item  $B_{l, 2, 1}^{\varphi_k, +}, l = 1, 2$ can be written as oscillatory integrals,
 \beq 
\begin{gathered}
B_{l, 2, 1}^{\varphi_k, +} f(x)   =  (2\pi)^{-3} \int_{\mbr^3 }   \int_{\mbr^3 } e^{\imath(x - y)\cdot \xi} e^{-\imath (T+\eps)|\xi|} \tilde b_{l, 2, 1}^{\varphi_k, +}(x, \xi) f(y)dy d\xi, 
\end{gathered}
\eeq
where $\tilde b_{l, 2, 1}^{\varphi_k, +}$ are symbols of order $-2-(l-1)$. The principal symbols of $B_{l, 2, 1}^{\varphi_k, +}$ in this representation  are given by 
\beq
\begin{gathered}
\sigma_{-2}(B_{1, 2, 1}^{\varphi_k, +})(x, y, \xi) =  -\frac{\pi^2}{(T+\eps)}  \phi_\eps(x) \varphi_k(\xi/|\xi|) |\xi|^{-2}, \\ 
 \sigma_{-3}(B_{2, 2, 1}^{\varphi_k, +})(x, y, \xi) =   -\frac{\pi^2}{\imath (T+\eps)}  \phi_\eps(x)  \varphi_k(\xi/|\xi|) |\xi|^{-3}
\end{gathered}
\eeq  
for $|\xi|$ large. 

\item $B_{l, 2, 2}^{\varphi_k, +}, l = 1, 2$ can be written as oscillatory integrals,
 \beq 
\begin{gathered}
B_{l, 2, 2}^{\varphi_k, +} f(x)   =  (2\pi)^{-3} \int_{\mbr^3 }   \int_{\mbr^3 } e^{\imath(x - y)\cdot \xi} e^{\imath (T-\eps)|\xi|} \tilde b_{l, 2, 1}^{\varphi_k, +}(x, \xi) f(y)dy d\xi, 
\end{gathered}
\eeq
where $\tilde b_{l, 2, 2}^{\varphi_k, +}$ are symbols of order $-2-(l-1)$. The  principal symbols of $B_{l, 2, 2}^{\varphi_k, +}$ in this representation are given by 
\beq
\begin{gathered}
\sigma_{-2}(B_{1, 2, 2}^{\varphi_k, +})(x, y, \xi) =  -\frac{\pi^2}{\eps} \phi_\eps(x)  \varphi_k(\xi/|\xi|) |\xi|^{-2}, \\ 
 \sigma_{-3}(B_{2, 2, 2}^{\varphi_k, +})(x, y, \xi) =   -\frac{\pi^2}{\imath \eps} \phi_\eps(x)  \varphi_k(\xi/|\xi|)|\xi|^{-3}
\end{gathered}
\eeq
for $|\xi|$ large. 

\end{enumerate}
\end{lemma}

\begin{lemma}\label{lm-b34new}
For $k = 1, 2, 3,$ and $l = 1, 2$, we can write 
\beq
I^{\varphi_k}L_\eps E_l^- = B_{l, 1}^{\varphi_k, -} + B_{l, 2, 1}^{\varphi_k, -} + B_{l, 2, 2}^{\varphi_k, -} + R_l^{\varphi_k, -}, 
\eeq
where $R_l^{\varphi_k, -}$ are regularizing operators and the rest are elliptic Fourier integral operators 
\beq
\begin{gathered}
B_{l, 1}^{\varphi_k, -}  \in I^{-1 -(l-1)}(\mbr^3\times \mbr^3; C'_{-T-\eps}), \\
B_{l, 2, 1}^{\varphi_k, -}  \in I^{-2-(l-1)}(\mbr^3\times \mbr^3; C'_{-T + \eps}), \quad B_{l, 2, 2}^{\varphi_k, -} \in I^{-2-(l-1)}(\mbr^3\times \mbr^3; C'_{T + \eps}). 
\end{gathered}
\eeq
Moreover,  
\begin{enumerate}
\item $B_{l, 1}^{\varphi_k, -}, l = 1, 2$ can be written as oscillatory integrals,
 \beq 
\begin{gathered}
B_{l, 1}^{\varphi_k, -} f(x)   =  (2\pi)^{-3} \int_{\mbr^3 }   \int_{\mbr^3 } e^{\imath(x - y)\cdot \xi} e^{-\imath (T+\eps)|\xi|} \tilde b_{l, 1}^{\varphi_k, -}(x, \xi) f(y)dy d\xi,   
\end{gathered}
\eeq
where $\tilde b_{l, 1}^{\varphi_k, -}$ are symbols of order $-1-(l-1)$. The principal symbols of $B_{l, 1}^{\varphi_k, -}$ in this representation  are given by 
\beq
\begin{gathered}
\sigma_{-1}(B_{1, 1}^{\varphi_k, -})(x, y, \xi) =  (-2\pi^2\imath \ln\frac{\eps}{T+\eps})  |\xi|^{-1} \varphi_k(\xi/|\xi|)\phi_\eps(x), \\
 \sigma_{-2}(B_{2, 1}^{\varphi_k, -})(x, y, \xi) = (2\pi^2\ln\frac{\eps}{T+\eps})   |\xi|^{-2} \varphi_k(\xi/|\xi|)\phi_\eps(x),
\end{gathered}
\eeq
for $|\xi|$ large. 
\item  $B_{l, 2, 1}^{\varphi_k, -}, l = 1, 2$ can be written as oscillatory integrals, 
 \beq 
\begin{gathered}
B_{l, 2, 1}^{\varphi_k, -} f(x)   =  (2\pi)^{-3} \int_{\mbr^3 }   \int_{\mbr^3 } e^{\imath(x - y)\cdot \xi} e^{\imath (-T+\eps)|\xi|} \tilde b_{l, 2, 1}^{\varphi_k, -}(x, \xi) f(y)dy d\xi, 
\end{gathered}
\eeq
where $\tilde b_{l, 2, 1}^{\varphi_k, -}$ are symbols of order $-2-(l-1)$. The principal symbols of $B_{l, 2, 1}^{\varphi_k, -}$ in this representation  are given by 
\beq
\begin{gathered}
\sigma_{-2}(B_{1, 2, 1}^{\varphi_k, -})(x, y, \xi) =   -\frac{\pi^2}{(T+\eps)} \phi_\eps(x) \varphi_k(-\xi/|\xi|)|\xi|^{-2},  \\ 
 \sigma_{-3}(B_{2, 2, 1}^{\varphi_k, -})(x, y, \xi) =    -\frac{\pi^2 \imath }{(T+\eps)} \phi_\eps(x)\varphi_k(-\xi/|\xi|)|\xi|^{-3},
\end{gathered}
\eeq  
for $|\xi|$ large. 

\item $B_{l, 2, 2}^{\varphi_k, -}, l = 1, 2$ can be written as oscillatory integrals, 
 \beq 
\begin{gathered}
B_{l, 2, 2}^{\varphi_k, -} f(x)   =  (2\pi)^{-3} \int_{\mbr^3 }   \int_{\mbr^3 } e^{\imath(x - y)\cdot \xi} e^{\imath (T+\eps)|\xi|} \tilde b_{l, 2, 1}^{\varphi_k, -}(x, \xi) f(y)dy d\xi, 
\end{gathered}
\eeq
where $\tilde b_{l, 2, 2}^{\varphi_k, -}$ are symbols of order $-2-(l-1)$. The  principal symbols of $B_{l, 2, 2}^{\varphi_k, -}$ in this representation are given by 
\beq
\begin{gathered}
\sigma_{-2}(B_{1, 2, 2}^{\varphi_k, -})(x, y, \xi) =  \frac{\pi^2}{\eps}  \phi_\eps(x)  \varphi_k(-\xi/|\xi|)|\xi|^{-2}, \\ 
 \sigma_{-3}(B_{2, 2, 2}^{\varphi_k, -})(x, y, \xi) =   \frac{\pi^2 \imath }{\eps}  \phi_\eps(x)\varphi_k(-\xi/|\xi|) |\xi|^{-3},
\end{gathered}
\eeq
for $|\xi|$ large. 
\end{enumerate}
\end{lemma}

Using Lemma \ref{lm-b12new} and \ref{lm-b34new}, we can write for $k = 1, 2, 3$ that 
\beqq\label{eq-ILk}
\begin{gathered}
  I^{\varphi_k} L_\eps u  
=  \tilde B_{1}^{\varphi_k, +} f_1 + \tilde B_{2}^{\varphi_k, +} f_2 
+  B_{1, 2, 2}^{\varphi_k, +} f_1 + B_{2, 2, 2}^{\varphi_k, +} f_2\\
+  \tilde B_{1}^{\varphi_k, -} f_1 + \tilde B_{2}^{\varphi_k, -} f_2
+  B_{1, 2, 1}^{\varphi_k, -} f_1 + B_{2, 2, 1}^{\varphi_k, -} f_2 
+ R_1^{\varphi_k} f_1 + R_2^{\varphi_k} f_2, 
\end{gathered}
\eeqq
where $R_1^{\varphi_k} = R_1^{\varphi_k, +} + R_1^{\varphi_k, -}, R_2^{\varphi_k} = R_2^{\varphi_k, +} + R_2^{\varphi_k, -}$ are regularizing operators and 
\beqq\label{eq-oplist}
\begin{gathered}
\tilde B_1^{\varphi_k, +} =  B_{1, 1}^{\varphi_k, +}  + B_{1, 2, 2}^{\varphi_k, -} \in I^{-1}(\mbr^3\times \mbr^3; C'_{T+\eps}),\\
\tilde B_2^{\varphi_k, +}  =   B_{2, 1}^{\varphi_k, +} + B_{2, 2, 2}^{\varphi_k, -} \in I^{-2}(\mbr^3\times \mbr^3; C'_{T+\eps}),  \\
\tilde B_1^{\varphi_k, -} =  B_{1, 1}^{\varphi_k, -}   + B_{1, 2, 1}^{\varphi_k, +} \in I^{-1}(\mbr^3\times \mbr^3; C'_{-T-\eps}), \\
  \tilde B_2^{\varphi_k, -}  =   B_{2, 1}^{\varphi_k, -} + B_{2, 2, 1}^{\varphi_k, +} \in I^{-2}(\mbr^3\times \mbr^3; C'_{-T-\eps}). 
\end{gathered}
\eeqq
Also, $B_{1, 2, 2}^{\varphi_k, +} \in I^{-2}(\mbr^3\times \mbr^3; C'_{T-\eps}), B_{2, 2, 2}^{\varphi_k, +}\in I^{-3}(\mbr^3\times \mbr^3; C'_{T-\eps})$ and $B_{1, 2, 1}^{\varphi_k, -} \in I^{-2}(\mbr^3\times \mbr^3; C'_{-T+\eps}), B_{2, 2, 1}^{\varphi_k, -}  \in I^{-3}(\mbr^3\times \mbr^3; C'_{-T+\eps})$. We note again that the orders of these FIOs acting on $f_1, f_2$ are lower than those corresponding terms in \eqref{eq-oplist}, even though the canonical relations are different.  
We sum the equations \eqref{eq-ILk} in $k$ to get 
\beqq\label{eq-ILnew}
\begin{gathered}
I^{\varphi} L_\eps u  
=  \tilde B_{1}^{\varphi, +} f_1 + \tilde B_{2}^{\varphi, +} f_2 
+  B_{1, 2, 2}^{\varphi, +} f_1 + B_{2, 2, 2}^{\varphi, +} f_2\\
+  \tilde B_{1}^{\varphi, -} f_1 + \tilde B_{2}^{\varphi, -} f_2
+  B_{1, 2, 1}^{\varphi, -} f_1 + B_{2, 2, 1}^{\varphi, -} f_2 
+ R_1^{\varphi} f_1 + R_2^{\varphi} f_2, 
\end{gathered}
\eeqq
where we let $\varphi = \sum_{k = 1}^3 \varphi_k$, and the terms are described as follows: 
\begin{enumerate}
\item $I^\varphi L_\eps u = \sum_{k = 1}^3  I^{\varphi_k} L_\eps u$. 

\item $\tilde B_{1}^{\varphi, +} = \sum_{k = 1}^3 \tilde B_{1}^{\varphi_k, +} \in I^{-1}(\mbr^3\times \mbr^3; C'_{T+\eps})$, $\tilde B_{2}^{\varphi_k, +} = \sum_{k = 1}^3 \tilde B_{2}^{\varphi_k, +} \in I^{-2}(\mbr^3\times \mbr^3; C'_{T+\eps}).$ Moreover, these operators can be expressed as in Lemma \ref{lm-b12new} and \ref{lm-b34new}.  The principal symbols of the operators are given by 
\beq
\begin{gathered}
\sigma_{-1}(\tilde B_{1}^{\varphi, +})(x, y, \xi) = (-2\pi^2\imath \ln\frac{\eps}{T+\eps})  \phi_\eps(x) \varphi(-\xi/|\xi|) |\xi|^{-1}, \\
 \sigma_{-2}(\tilde B_{2}^{\varphi, +})(x, y, \xi) = (-2\pi^2  \ln\frac{\eps}{T+\eps}) \phi_\eps(x) \varphi(-\xi/|\xi|)   |\xi|^{-2}, 
\end{gathered}
\eeq
for $|\xi|$ large. 
\item $\tilde B_1^{\varphi, -} = \sum_{k = 1}^3 \tilde B_1^{\varphi_k, -} \in I^{-1}(\mbr^3\times \mbr^3; C'_{-T-\eps}),$ $ \tilde B_2^{\varphi, -}  = \sum_{k = 1}^3  \tilde B_2^{\varphi_k, -} \in I^{-2}(\mbr^3\times \mbr^3; C'_{-T-\eps}).$
Moreover, these operators can be expressed as in Lemmas \ref{lm-b12new} and \ref{lm-b34new}. The principal symbols of the operators are given by 
\beq
\begin{gathered}
\sigma_{-1}(\tilde B_{1}^{\varphi, -})(x, y, \xi) =   (-2\pi^2\imath \ln\frac{\eps}{T+\eps})  |\xi|^{-1} \varphi(\xi/|\xi|) \phi_\eps(x),\\
 \sigma_{-2}(\tilde B_{2}^{\varphi, -})(x, y, \xi) =  (2\pi^2  \ln\frac{\eps}{T+\eps})   |\xi|^{-2}  \varphi(\xi/|\xi|) \phi_\eps(x), 
\end{gathered}
\eeq
for $|\xi|$ large. 
\item $B_{1, 2, 2}^{\varphi, +} = \sum_{k = 1}^3 B_{1, 2, 2}^{\varphi_k, +} \in I^{-2}(\mbr^3\times \mbr^3; C'_{T-\eps})$,  $B_{2, 2, 2}^{\varphi, +} = \sum_{k = 1}^3 B_{2, 2, 2}^{\varphi_k, +}\in I^{-3}(\mbr^3\times \mbr^3; C'_{T-\eps})$, $B_{1, 2, 1}^{\varphi, -} = \sum_{k = 1}^3 B_{1, 2, 1}^{\varphi_k, -} \in I^{-2}(\mbr^3\times \mbr^3; C'_{-T+\eps}),$  $B_{2, 2, 1}^{\varphi, -}  = \sum_{k = 1}^3 B_{2, 2, 1}^{\varphi_k, -}  \in I^{-3}(\mbr^3\times \mbr^3; C'_{-T+\eps})$.

\item $R_l^{\varphi} = \sum_{k = 1}^3 R_l^{\varphi_k}, l=1, 2$ are regularizing operators.  \\
\end{enumerate}

Now we are ready to solve for $f_1, f_2$ from \eqref{eq-IL} and \eqref{eq-ILnew} modulo more regular terms. Assume that $f_1, f_2$ are supported in $\mcr_{T-1+2\eps, T+1-2\eps}$ for $\eps>0$ small. 
First, we can find a parametrix $Q^+ \in I^{1}(\mbr^3\times \mbr^3; C_{-T-\eps})$ such that $Q^+ \circ \tilde B_{1}^+ = \id$ on $\mcr_{T- 1 + 2\eps, T+1-2\eps}$ modulo a regularizing operator.  Here, we used that $\phi_\eps = 1$ on $\mcb_{1-2\eps}$. Applying $Q^+$ to \eqref{eq-IL}, we get 
\beqq\label{eq-q1} 
\begin{gathered}
 Q^+ I L_\eps u  
 =  f_1 + Q^+ \tilde B_{1}^- f_1  + Q^+ \tilde B_{2}^+ f_2  + Q^+ \tilde B_{2}^- f_2  + \mci^{-1}(f_1) +\mci^{-2}(f_2), 
\end{gathered}
\eeqq
where we have denoted by $\mci^{m}(f)$ a collection of FIOs acting on $f$ that are of order at most $m$ and with canonical relations of the type $C_a$ with some $a\in \mbr.$  The specific forms of these operators will not be important for the analysis below. 
Next, we find a parametrix $Q^{\varphi, +} \in I^{1}(\mbr^3\times \mbr^3; C_{-T-\eps})$ such that $Q^{\varphi, +} \circ \tilde B_{\varphi, 1}^+ = \id$ on $\mcr_{T- 1 + 2\eps, T+1-2\eps}$ modulo a regularizing operator.  Applying $Q^{\varphi, +}$ to \eqref{eq-ILnew}, we get 
\beqq\label{eq-q2}
\begin{gathered}
Q^{\varphi, +} I^{\varphi} L_\eps u  
 = f_1 +   Q^{\varphi, +} \tilde B_{1}^{\varphi, -} f_1  
 + Q^{\varphi, +} \tilde B_{2}^{\varphi, +} f_2  
   + Q^{\varphi, +} \tilde B_{2}^{\varphi, -} f_2  +\mci^{-1}(f_1) +\mci^{-2}(f_2). 
\end{gathered}
\eeqq
From \eqref{eq-q1} and \eqref{eq-q2}, we get 
\beqq\label{eq-q3} 
\begin{gathered}
 Q^+ I L_\eps u  - Q^{\varphi, +} I^{\varphi} L_\eps u 
 =   (Q^+ \tilde B_{1}^- - Q^{\varphi, +} \tilde B_{1}^{\varphi, -} )f_1  \\
 + (Q^+ \tilde B_{2}^+ - Q^{\varphi, +} \tilde B_{2}^{\varphi, +} )f_2  + (Q^+ \tilde B_{2}^- - Q^{\varphi, +} \tilde B_{2}^{\varphi, -} )f_2  + \mci^{-1}(f_1) +\mci^{-2}(f_2). 
\end{gathered}
\eeqq
We note that $Q^+ \tilde B_{1}^- - Q^{\varphi, +} \tilde B_{1}^{\varphi, -} \in I^{0}(\mbr^3\times \mbr^3; C'_{-2T-2\eps})$, and it can  be written as an oscillatory integral 
 \beqq\label{eq-symcal}
\begin{gathered}
(Q^+ \tilde B_{1}^- - Q^{\varphi, +} \tilde B_{1}^{\varphi, -}) f(z)   = (2\pi)^{-3} \int_{\mbr^3 }   \int_{\mbr^3 } e^{\imath(z - y)\cdot \xi} e^{\imath (-2T-2\eps)|\xi|} a(y, \xi)f(y)dy d\xi, 
\end{gathered}
\eeqq
where the principle symbol (the leading order term of $a$) is 
\beq
\sigma_0(Q^+ \tilde B_{1}^- - Q^{\varphi, +} \tilde B_{1}^{\varphi, -})(y, \xi) = (1 - \varphi(\xi/|\xi|)/\varphi(-\xi/|\xi|)) 
\eeq
for $y + (T+\eps)\xi/|\xi|\in \mcb_{1-2\eps}$. Note that the principal symbol is non-vanishing. Next, $Q^+ \tilde B_{2}^+ - Q^{\varphi, +} \tilde B_{2}^{\varphi, +} \in I^{-1}(\mbr^3\times \mbr^3; C'_{0})$ but the principal symbol vanishes so the term actually belongs to $\mci^{-2}(f_2)$. Finally, 
$ Q^+ \tilde B_{2}^- - Q^{\varphi,+} \tilde B_{2}^{\varphi, -} \in I^{-1}(\mbr^3\times \mbr^3; C'_{-2T-2\eps})$, and a similar calculation as \eqref{eq-symcal} shows that the principal symbol is 
 \beq
\sigma_{-1} (Q^+ \tilde B_{2}^- - Q^{\varphi,+} \tilde B_{2}^{\varphi, -} )(y, \xi) = \imath(1 - \varphi(\xi/|\xi|)/\varphi(-\xi/|\xi|))|\xi|^{-1}
\eeq
for $y + (T+\eps)\xi/|\xi|\in \mcb_{1-2\eps}$. Now we let $W^+\in I^{0}(\mbr^3\times \mbr^3; C'_{2T+2\eps})$ be a parametrix of $Q^+ \tilde B_{2}^+ - Q^{\varphi, +} \tilde B_{2}^{\varphi, +}$ and we get from \eqref{eq-q3} that 
\beqq\label{eq-q4}
\begin{gathered}
W^+ Q^+ I L_\eps u  - W^+ Q^{\varphi, +} I^{\varphi} L_\eps u 
 =   f_1    +  U^+ f_2  + \mci^{-1}(f_1) +\mci^{-2}(f_2), 
\end{gathered}
\eeqq
where $U^+$ is a pseudo-differential operator of order $-1$ with principal symbol $\imath |\xi|^{-1}$ on $\mcr_{T-1+2\eps, T+1-2\eps}$.

Next we find a parametrix $Q^- \in I^{2}(\mbr^3\times \mbr^3; C_{T+\eps})$ such that $Q^- \circ \tilde B_{2}^- = \id$ on $\mcr_{T- 1 + 2\eps, T+1-2\eps}$ modulo a regularizing operator.  Applying $Q^-$ to \eqref{eq-IL}, we get 
\beqq\label{eq-q5} 
\begin{gathered}
 Q^- I L_\eps u  
 =  f_1 + Q^- \tilde B_{1}^+ f_1  + Q^- \tilde B_{2}^+ f_2  + Q^- \tilde B_{2}^- f_2  + \mci^{-1}(f_1) +\mci^{-2}(f_2). 
\end{gathered}
\eeqq
Now we find a parametrix $Q^{\varphi, -} \in I^{1}(\mbr^3\times \mbr^3; C_{T+\eps})$ such that $Q^{\varphi, -} \circ \tilde B_{\varphi, 1}^- = \id$ on $\mcr_{T- 1 + 2\eps, T+1-2\eps}$ modulo a regularizing operator.  Applying $Q^{\varphi, -}$ to \eqref{eq-ILnew}, we get 
\beqq\label{eq-q6}
\begin{gathered}
Q^{\varphi, -} I^{\varphi} L_\eps u  
 = f_1 +   Q^{\varphi, -} \tilde B_{1}^{\varphi, +} f_1  
 + Q^{\varphi, -} \tilde B_{2}^{\varphi, +} f_2  
   + Q^{\varphi, -} \tilde B_{2}^{\varphi, -} f_2  +\mci^{-1}(f_1) +\mci^{-2}(f_2). 
\end{gathered}
\eeqq
From \eqref{eq-q5} and \eqref{eq-q6}, we get 
\beqq\label{eq-q7} 
\begin{gathered}
 Q^- I L_\eps u  - Q^{\varphi, -} I^{\varphi} L_\eps u 
 =   (Q^- \tilde B_{1}^+ - Q^{\varphi, -} \tilde B_{1}^{\varphi, +} )f_1  \\
 + (Q^- \tilde B_{2}^+ - Q^{\varphi, -} \tilde B_{2}^{\varphi, +} )f_2  + (Q^- \tilde B_{2}^- - Q^{\varphi, -} \tilde B_{2}^{\varphi, -} )f_2  + \mci^{-1}(f_1) +\mci^{-2}(f_2). 
\end{gathered}
\eeqq
We note that $Q^- \tilde B_{1}^+ - Q^{\varphi, -} \tilde B_{1}^{\varphi, +} \in I^{0}(\mbr^3\times \mbr^3; C'_{2T+2\eps})$. Writing the operator as in \eqref{eq-symcal}, we find that the principal symbol is 
\beq
\sigma_0( Q^- \tilde B_{1}^+ - Q^{\varphi, -} \tilde B_{1}^{\varphi, +}  )(y, \xi) = 1 - \varphi(\xi/|\xi|)/\varphi(-\xi/|\xi|)
\eeq
for $y - (T+\eps)\xi/|\xi|\in \mcb_{1-2\eps}$.  Also, $Q^- \tilde B_{2}^- - Q^{\varphi, -} \tilde B_{2}^{\varphi, -} \in I^{-1}(\mbr^3\times \mbr^3; C'_{0})$ but the principal symbol is $0$ so the term belongs to $\mci^{-2}(f_2)$. Finally, $ Q^- \tilde B_{2}^+ - Q^{\varphi,-} \tilde B_{2}^{\varphi, +} \in I^{-1}(\mbr^3\times \mbr^3; C'_{2T+2\eps})$ and the principal symbol is 
 \beq
\sigma_{-1}(Q^- \tilde B_{2}^+ - Q^{\varphi,-} \tilde B_{2}^{\varphi, +})(y, \xi) = -\imath (1 - \varphi(\xi/|\xi|)/\varphi(-\xi/|\xi|))|\xi|^{-1}
\eeq
for $y - (T+\eps)\xi/|\xi|\in \mcb_{1-2\eps}$.  Now we let $W^-\in I^{0}(\mbr^3\times \mbr^3; C'_{2T+2\eps})$ be a parametrix of $Q^- \tilde B_{2}^+ - Q^{\varphi, -} \tilde B_{2}^{\varphi, +}$ and we get from \eqref{eq-q7} that 
\beqq\label{eq-q8}
\begin{gathered}
W^- Q^- I L_\eps u  - W^- Q^{\varphi, -} I^{\varphi} L_\eps u 
 =   f_1    +  U^- f_2  + \mci^{-1}(f_1) +\mci^{-2}(f_2), 
\end{gathered}
\eeqq
where $U^-$ is a pseudo-differential operator of order $-1$ with principal symbol $-\imath|\xi|^{-1}$.

Finally, from \eqref{eq-q4} and \eqref{eq-q8}, we get 
\beqq\label{eq-q9}
\begin{gathered}
W^+ Q^+ I L_\eps u  - W^+ Q^{\varphi, +} I^{\varphi} L_\eps u - W^- Q^- I L_\eps u  + W^- Q^{\varphi, -} I^{\varphi} L_\eps u \\
 =  ( U^+ - U^-) f_2  +\mci^{-1}(f_1) +\mci^{-2}(f_2). 
\end{gathered}
\eeqq
Because $U^+ - U^-$ is a pseudo-differential operator of order $-1$ with symbol $2\imath |\xi|^{-1}$. We can find a parametrix $S$ which is a pseudo-differential operator of order $1$ with symbol $(2\imath)^{-1}|\xi|$. Thus we get from \eqref{eq-q9} that 
 \beqq\label{eq-q10}
\begin{gathered}
  f_2  = SW^+ Q^+ I L_\eps u  - SW^+ Q^{\varphi, +} I^{\varphi} L_\eps u - SW^- Q^- I L_\eps u  \\
  + SW^- Q^{\varphi, -} I^{\varphi} L_\eps u + \mci^0(f_1) +  \mci^{-1}(f_2). 
\end{gathered}
\eeqq
Then we can use \eqref{eq-q8} and \eqref{eq-q10} to solve for $f_1$ as
 \beqq\label{eq-q11}
\begin{gathered}
f_1 = W^- Q^- I L_\eps u  - W^- Q^{\varphi, -} I^{\varphi} L_\eps u 
     -  U^- f_2  + \mci^{-1}(f_1) +\mci^{-2}(f_2)\\
      =  W^- Q^- I L_\eps u  - W^- Q^{\varphi, -} I^{\varphi} L_\eps u\\
       -  U^- (SW^+ Q^+ I L_\eps u  - SW^+ Q^{\varphi, +} I^{\varphi} L_\eps u - SW^- Q^- I L_\eps u) 
      + \mci^{-1}(f_1) +\mci^{-2}(f_2). 
\end{gathered}
\eeqq

This concludes the microlocal inversion. In particular, we have solved $f_1, f_2$ up to more regular terms. We remark that it is possible to use \eqref{eq-q10} and \eqref{eq-q11} to analyze what singularities (wave front sets) of $f_1, f_2$ can be reconstructed from $L_\eps u$, even if $f_1, f_2$ are not compactly supported in $\mcr_{T-1 + 2\eps, T+1-2\eps}$. But we will not pursue it here.

\section{Proof of Theorem \ref{thm-main}} \label{sec-pf}
We start with the Sobolev estimates for recovering $f_1, f_2$. We recall the mapping properties of FIOs of graph type on Sobolev spaces, see \cite[Section 25.3]{Ho4}. In particular, for $A\in I^\mu(\mbr^3\times \mbr^3; C'_a),$ where $a\in \mbr$, $\mu\in \mbr$, we have for $s\in \mbr$ that 
\beq
\|A f\|_{H^{s}(\mbr^3)} \leq C \|f\|_{H^{s+\mu}(\mbr^3)}
\eeq 
for some $C>0$. Hereafter, we use $C$ for a generic constant which could depend on $\eps$ but is uniform in $f_1, f_2$. 
In \eqref{eq-q10}, we recall that $S\in \Psi^{-1}(\mbr^3), W^\pm \in \mci^0$, and $Q^{\varphi, \pm}, Q^{\pm} \in \mci^1$. We get from \eqref{eq-q10} that 
\beq
\|f_2\|_{H^s} \leq C(\|IL_\eps u\|_{H^s} + \|I^\varphi L_\eps u\|_{H^s}) + C \|f_1\|_{H^s} + C\|f_2\|_{H^{s-1}}. 
\eeq
Similarly, in \eqref{eq-q11}, we know that $U^\pm \in \Psi^{-1}(\mbr^3)$. We get 
\beq
\|f_1\|_{H^s} \leq C(\|I L_\eps u\|_{H^{s+1}} + \|I^\varphi L_\eps u\|_{H^{s+1}} ) + C\|f_1\|_{H^{s-1}} + C\|f_2\|_{H^{s-2}}. 
\eeq 
From these two estimates,  we obtain that 
\beq
\|f_1\|_{H^{s+1}} + \|f_2\|_{H^{s}} \leq C (\|I L_\eps u\|_{H^{s+2}} + \|I^\varphi L_\eps u\|_{H^{s+2}} ) + C\|f_1\|_{H^{s}} + C\|f_2\|_{H^{s-1}}. 
\eeq
We observe that for $s + 2 \geq 0$ an integer, 
\beq
\|I L_\eps u\|_{H^{s+2}} \leq C \|L u\|_{H^{s+2}}, \quad \|I^\varphi L_\eps u\|_{H^{s+2}} \leq C \|L u\|_{H^{s+2}}, 
\eeq
which can be seen by directly estimating the integral in $I$ and $I^\varphi.$ 
Thus 
\beqq\label{eq-sob}
\|f_1\|_{H^{s+1}} + \|f_2\|_{H^{s}} \leq C \|L u\|_{H^{s+2}}   + C\|f_1\|_{H^{s}} + C\|f_2\|_{H^{s-1}}. 
\eeqq

Next, we remove the last two terms. We will need the following uniqueness result for the light ray transform with partial data. 
 \begin{lemma}\label{lm-inj}
 Let $u\in H^s(\mcm), s\geq 0$ and $Lu = 0$. Then $u = 0$ on the set 
 \beqq\label{eq-bcone}
 \{(t, x)\in \mcm :  T-t \leq |x| < T+ 1 - t \}.
 \eeqq
\end{lemma}
\bpf
Let $\phi(x)$ be the characteristic function of $\mcb$ in $\mbr^3$. Let $\chi(t)$ be the characteristic function of $[0, T]$ in $\mbr.$ We write  
\beq 
Lu(y, v) = \int_{\mbr}\chi(t)\phi(y)u(t, y+ tv - Tv) dt, 
\eeq
where $y\in \mbr^3, v\in \mbs^2.$ We follow the proof of the Fourier Slice Theorem to compute that 
\beqq\label{eq-A}
\begin{gathered}
0 = \int_{\mbr^3} e^{-\imath y \cdot \xi} Lu(y, v) dy = \int_{\mbr^3} e^{-\imath y \cdot \xi} \chi(t)\phi(y)u(t, y+ tv - Tv) dt dy\\
 =  \int_{\mbr^3} e^{-\imath (x - tv + Tv) \cdot \xi} \chi(t)\phi(x - tv + Tv)u(t, x) dt dx = e^{-\imath Tv\cdot \xi} \hat A(-v\cdot \xi, \xi), 
\end{gathered}
\eeqq
where $\hat A(\tau, \xi)$ is the Fourier transform of $A(t, x) = \chi(t)\phi(x - tv + Tv)u(t, x)$. Note that $A$ is compactly supported in $\mbr^{3+1}$ so $\hat A$ is analytic. We see from \eqref{eq-A} that $\hat A$ vanishes on some open set with non-empty interior. Thus $\hat A$ is identically zero. So we get  
$ \chi(t)\phi(x - tv + Tv)u(t, x) =0$ so $u$ vanishes on the set \eqref{eq-bcone}. 
\epf

\bpf[Proof of Theorem \ref{thm-main}]
Let $f_1 \in H^{s+1}, f_2\in H^{s}, s\geq 0$ be compactly supported in $\mcr_{T-1, T+1}$. Then there exists $\eps>0$ such that \eqref{eq-sob} holds with $C$ depending on $\eps$. We prove \eqref{eq-mainest}, that is, 
\beq
\|f_1\|_{H^{s+1}} + \|f_2\|_{H^s} \leq C \|Lu\|_{H^{s+2}}.
\eeq
By contradiction, we assume that for $n = 1, 2, \cdots,$ there are $f_1^n \in H^{s+1}, f_2^n \in H^{s}$ compactly supported in $\mcr_{T-1, T+1}$ such that $\|f_1\|_{H^{s+1}} = 1, \|f_2\|_{H^s} =1$, and 
\beqq\label{eq-con}
\|f_1^n\|_{H^{s+1}} + \|f_2^n\|_{H^s} \geq n \|Lu^n\|_{H^{s+2}}, 
\eeqq
where $u^n$ is the solution of \eqref{eq-cau} with Cauchy data $f_1^n, f_2^n.$ First, we note that on a fixed compact set, $H^{s+1}\times H^s$ is compactly embedded in $H^{s}\times H^{s-1}$. Thus by passing to a subsequence, we can assume that $(f_1^n, f_2^n)$ converges in $H^s\times H^{s-1}$, and it follows from \eqref{eq-sob} that $(f_1^n, f_2^n)$ converges to some $(f_1, f_2)$ in $H^{s+1}\times H^s.$ For the Cauchy problem \eqref{eq-cau}, we have the standard energy estimate
\beq
\|u\|_{H^{s+1}}\leq C(\|f_1\|_{H^{s+1}} + \|f_2\|_{H^s}).
\eeq
Therefore, $u^n$ converges to $u$ in $H^{s+1}$. We get from \eqref{eq-con} that $0 = \|Lu\|_{H^{s+2}} \geq \|Lu\|_{H^{s+1}}$. By Lemma \ref{lm-inj} and the finite speed of propagation for \eqref{eq-cau}, we deduce that $f_1 = f_2 = 0$. But this contradicts \eqref{eq-sob}. Thus, we proved \eqref{eq-mainest}, and this completes the proof of Theorem \ref{thm-main}. 
\epf

\section{The numerical experiment}\label{sec-num}

\subsection{The numerical setup} 
We conduct numerical experiments for a 2D universe. 
In Section \ref{sec-main}, we explained the local nature of the problem. So it suffices to consider the numerical simulation on a compact region. We take $\mcm = [0, 2]\times [-7, 7]^2$ and 
consider the Cauchy problem of the standard wave equations 
\beqq\label{eq-cau2d}
\begin{gathered}
\square u(t, x) = 0, \ \  t > 0, x\in [-7, 7]^2 \\
u(0, x) = f(x), \ \ \p_t u(0, x) = 0,
\end{gathered}
\eeqq
where $\square = \p_t^2 +   \lap $ with $\lap = -\sum_{i = 1}^2 \p_{x_i}^2.$  Note that we take $\p_t u = 0$ at $t = 0$ for simplicity. Also, instead of using techniques such as Perfectly Matched Layer, we will impose zero boundary conditions on the lateral boundary $[0, 2]\times \p[-7, 7]^2$. This means that waves are reflected from the boundary. In the numerical experiments, we will make sure that either these reflected waves are not observed (in the partial data reconstruction) or we restrict $f$ to be supported in a sufficiently small region (for the full data reconstruction) so that the waves do not meet the boundary for $t\in[0, 2]$. Hence, they do not affect the reconstruction problem.

Let $\mcu$ be an open set of $\{2\}\times [-7, 7]^2$. We consider light-like geodesics from $\{0\}\times [-7, 7]^2$ that meet $\mcu$ and the light ray transform 
\beqq\label{eq-lray}
Lu(y, v) = \int_{0}^{T}u(t, y+ tv - Tv) dt, \quad y\in \mcu, v\in \mbs^1. 
\eeqq 
This will be the simulated CMB data. The inverse problem is to recover $f$ from $Lu.$ In principle, we treat this as a PDE constrained optimization problem. However, for \eqref{eq-cau2d}, it is relatively straightfoward to find the (discretized) solution operator. We let $u = Sf$ and $w = Lu$ be the measurement. Then we can solve the more direct optimization problem, 
\beq
\min_{f} \|L S f - w\| + \alpha \|f\|
\eeq
with suitable norms and regularization terms. Here, $\alpha \geq 0$ is a regularization parameter.

We start with the discretization of $S$. We will solve \eqref{eq-cau2d} by the classical finite difference method on a rectangular grid. We refer to  \cite{langtangen2017finite} for the details. The mesh in time consists of $T$ time points $t_0 = 0 < t_1 <  \cdots < t_T,$ with constant spacing $\Delta t$, and mesh points in each space direction are given as
$$x_1 < x_2 < \cdots < x_n \quad \mbox{and} \quad y_1 < y_2 < \cdots < y_n$$ 
with equally spaced spatial grids $\Delta x$ and $\Delta y$.  For any time point $t_\tau$, we denote $u_{i,j}^\tau$ as $u$ at mesh point $(x_i,y_j,t_\tau)$ and let $\bfu^\tau \in \bbR^{n^2}$ be the discretized and vectorized image containing values $u_{i,j}^\tau$.  That is, at $\tau=0$ (i.e., the initial time point $t_0$), the unknown initial condition is given by $\bfu^0 = \bff$. For large 2D meshes, vectorization is essential for efficient computation; thus, we describe the solution operator in terms of matrix multiplications. Let $\bfT_x$ and $\bfT_y$ represent Poisson matrices obtained using finite differences, i.e.,
$$\bfT_x = \bfT_y =
\begin{bmatrix}
-2 & 1 & & & \\
1 & -2 & 1 &  &  \\
 & \ddots & \ddots & \ddots & \\
   & & 1 & -2 & 1\\
   & & & 1 & -2
\end{bmatrix} \in \bbR^{n \times n}.$$
Then, given the discretized initial condition $\bff$, the solution of the PDE at the first time point can be computed as
$$\bfu^1 = \bfT \bfu^0 \quad \mbox{
where} \quad \bfT = \bfI + \frac{\sigma^2}{2}(\bfI \otimes \bfT_x) + \frac{\gamma^2}{2}(\bfT_y \otimes \bfI)$$
with $\sigma = \gamma = \Delta t/ \Delta x$.
Subsequent time points can be computed as,
\begin{equation}
\bfu^\tau = 2 \bfT \bfu^{\tau-1} - \bfu^{\tau-2} \quad \mbox{for} \quad \tau = 2, \ldots, T.
\end{equation}
One can represent the entire vector of spatio-temporal voxels $\bfu \in \bbR^{n^2T}$ as 
\begin{equation*}
     \bfu = \begin{bmatrix} \bfu^1  \\ \bfu^2 \\ \vdots \\ \bfu^T\end{bmatrix}  = \bfS \bfu^0  = \bfS \bff,
\end{equation*}
for some matrix $\bfS \in \bbR^{n^2T \times n^2}.$  

For the discretization of the light ray transform $L$, we use the method described in \cite{COW}. Roughly speaking, we think of  each grid point on $\{0\}\times [-7, 7]^2$ as a point source, and each grid point on $\{2\}\times [-7, 7]^2$ as a receiver. Then we find the light rays from each source point to the (nearest) receivers which yields the discretization of the light rays. Assume there are $N$ source grid locations. Then for each source location $s_i$ for $i = 1, 2, \ldots, N$, there are $m_i$ observations that are collected in vector,
\begin{equation} \label{localRayTrace}
\bfb_i = \bfH_i \bfu + \mathbf{e}_i
\end{equation}
where $\bfb_i \in \bbR^{m_i}$ and $\bfH_i \in \bbR^{m_i \times n^2T}$ with noise contamination $\bfe_i \in \mathbb{R}^{m_i}$. The number of observations $m_i$ corresponds to the number of observable detector grid locations such that light-rays from source point $s_i$ have a non-zero interaction with the space-time mesh grid.  Thus, if we let
$$\bfb = \begin{bmatrix} \bfb_1 \\ \vdots \\ \bfb_N \end{bmatrix}\quad \mbox{and} \quad \bfH = \begin{bmatrix} \bfH_1 \\ \vdots \\ \bfH_N \end{bmatrix}, $$
we have an inverse problem where noisy data is given by 
\begin{equation} 
\label{eq:inverseprob}
 \bfb  = \bfH \bfu + \bfe  = \underbrace{\bfH \bfS}_{\bfA} \bff + \bfe,
\end{equation}
where $\bfA\in \bbR^{m \times n^2}$ with $m = \sum_{i=1}^N m_i$, and the goal of the reconstruction problem is to obtain an approximation of the initial time point $\bff$, given $\bfb$ and $\bfA$. We assume that $\bfe$ is i.i.d. standard Gaussian noise.

\subsection{The spectral analysis} 
Before we consider the reconstruction problem, we investigate some of the spectral properties of $\bfA$.  For the discretization of the world sheet, the $x,y$-values are linearly spaced on $[-7,7] \times [-7, 7]$ and discretized using $51 \times 51$ points.  Thus, $\bff \in \bbR^{2601}$.  We assume that we have $T = 40$ time slices where the planes are fixed to be the midpoints of the subintervals of $[0, 2]$ so that the first and last do not coincide with the source and detector planes. The PDE solution matrix $\bfS$ that takes the initial condition $\bff$ to the full space-time representation is $104,040 \times 2,601$. The cone beam light ray transform matrix $\bfH$ is defined by the locations of the sources and detectors. The sources and detectors are equally spaced on a $51 \times 51$ grid on $[-7, 7] \times [ -7, 7]$. For the full data reconstruction problem (where all detectors receive measurements), the resulting matrix $\bfH$ is $103,384 \times 104,040$. 

We also consider partial data reconstruction problems where the locations of the detectors remain fixed, but the active detectors that collect measurements are only those located in a square region around the origin. For different sizes of detector grids, we provide in Table \ref{tab:grid} the corresponding total number of observed measurements.  As expected, as the number of active detectors decreases, the total number of observations also decreases and the condition number of the forward model matrix $\bfA$ increases rapidly. The number of unknowns $n^2 = 2,601$ remains fixed. 
\begin{table}[bthp]
\begin{tabular}{c|c|c}
     detector grid & $m$ & $\kappa(\bfA)$\\ \hline
     $51 \times 51$ & 103,384 & 60.1800 \\
     $21 \times 21$ & 21,168 & 558.0165 \\
     $7 \times 7$ & 2,352 & $\infty$\\
     $3 \times 3$ & 432 & $\infty$\\
\end{tabular}
\caption{For different sized detector grids, we provide the total number of observations $m$, which corresponds to the number of rows of $\bfA$.  The number of columns of $\bfA$ remains fixed at $2,601$. $\kappa$ denotes the numerically computed condition number.}
\label{tab:grid}
\end{table}

\begin{figure}[bthp]
\includegraphics[width = .95\textwidth]{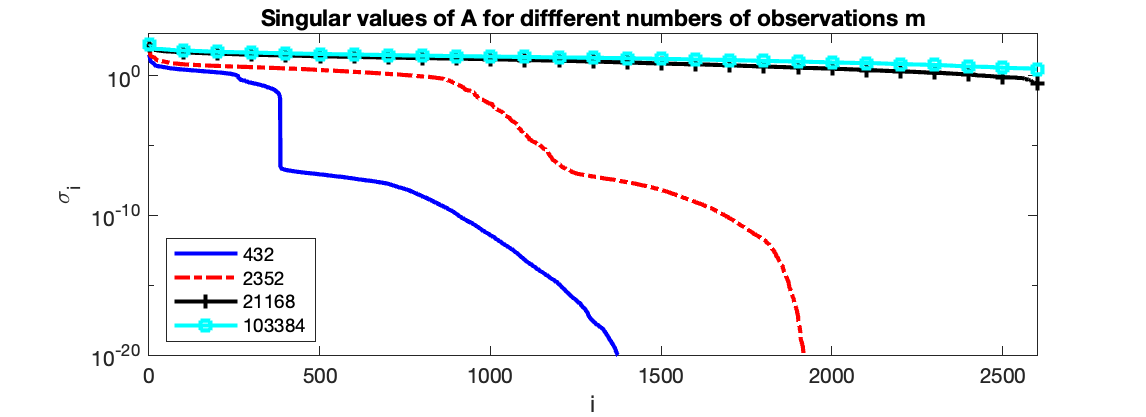}
\caption{Singular values $\sigma_i$ for matrices $\bfA$ corresponding to different sized detector grids. In the legend, $m$ corresponds to the number of observed measurements.}
\label{fig:spectrum}
\end{figure}

For each of the matrices $\bfA$ corresponding to different numbers of detectors, we provide the decay of the singular values in Figure \ref{fig:spectrum}.  We observe that with more detectors (e.g., a larger detector grid), there are very few small singular values and so the matrix is well-conditioned.  As the number of observations decreases, the problem becomes more ill-posed, evident in the small (and numerically zero) singular values.

\begin{figure}[bthp]
\includegraphics[width =.85\textwidth]{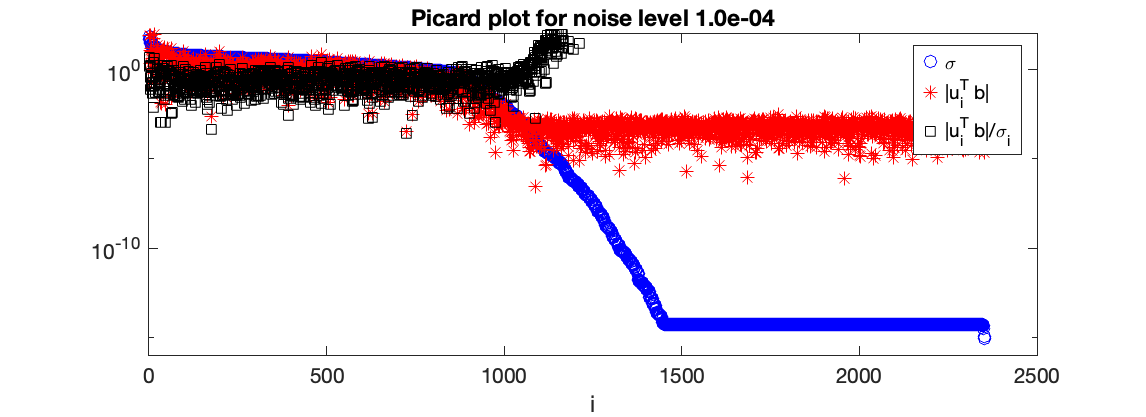}
\caption{Picard plot for the $7 \times 7$ detector grid demonstrating the impact of noise on the SVD and solution coefficients.}
\label{fig:Picard}
\end{figure}

For the $7 \times 7$ grid, we provide in Figure \ref{fig:Picard} the Picard plot \cite{hansen2010discrete}, which contains the singular values of $\bfA$, $\sigma_i$, the SVD coefficients $|\bfu_i\t \bfb|$ and the solution coefficients $\frac{|\bfu_i\t \bfb|}{\sigma_i}$ where $\bfu_i$ and $\bfv_i$ are the left and right singular vectors of $\bfA$ respectively.  The observation vector $\bfb$ was constructed using the true image in the bottom row of Figure \ref{fig:full_recon} with $10^{-4}$ noise level.  In addition to the decay of singular values, we observe that the SVD coefficients decay on average faster than the singular values and they level off at the noise level.  We also observe that once the SVD coefficients level off at the noise level, the solution coefficients increase significantly.  Thus, the inverse solution, $\sum_i \frac{\bfu_i\t \bfb}{\sigma_i} \bfv_i$, will likely be dominated by noise. Regularization will be essential to obtain reasonable reconstructions, especially for the partial data problem. Next we provide numerical results and comparisons for the full data and partial data reconstruction problems.

\subsection{Full data reconstruction problem}
\label{sub:full_recon}

For the full data problem set up, we have $51^2$ detectors and the corresponding matrix $\bfA$ has a condition number of $60.18$. This is relatively small, so we anticipate that even with noise in the data, solving 
the least-squares (LS) problem should yield good solutions.  So for the discrete inverse problem \eqref{eq:inverseprob}, we consider solving  
\begin{equation} \label{eq:LS}
\min_\bff \| \bfA \bff - \bfb \|^2,
\end{equation}
where unless specified otherwise, $\|\cdot\|$ will denote the Euclidean norm.

\begin{figure} 
\includegraphics[width = \textwidth]{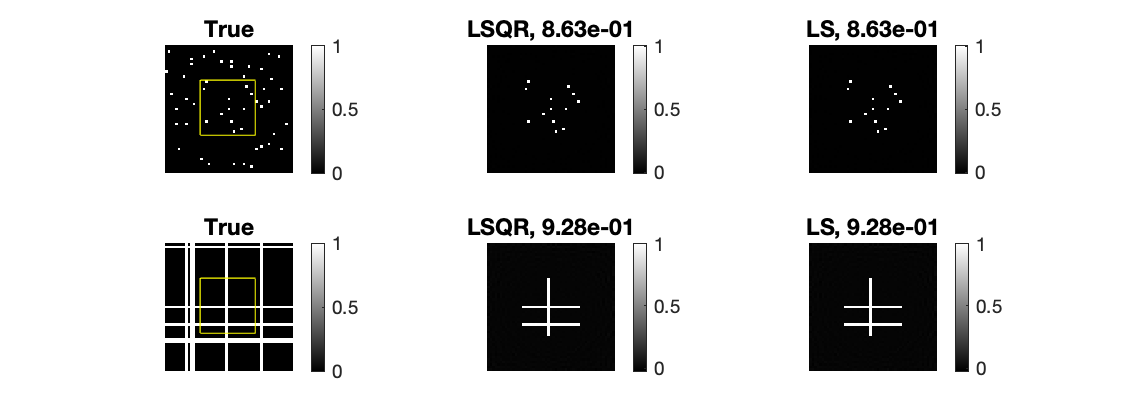}
\caption{We provide two example true images, where the box identifies the region $[-3,3] \times [-3,3]$ used for the full data problem. We provide the reconstructions for the full data problem using iterative method LSQR (after $100$ iterations) and the LS solution. Relative reconstruction error norms are provided in the titles.}
\label{fig:full_recon}
\end{figure}

We use two different true images, denoted $\bff_{\rm true}$, to represent the true initial condition, see the first column of Figure \ref{fig:full_recon}.  Each image is of size $51 \times 51$. The first one contains random dots that simulate gravitational wave point sources. The second image contains randomly placed vertical and horizontal lines that simulate sources of gravitational plane waves or cosmic strings. 
For the full-data reconstruction, we need to avoid the reflected waves from the lateral boundary and ensure that we collect all rays from the cone beam light transform that intersect the object. By the finite speed of propagation, it suffices to restrict the non-zero pixels of the true image to $[-3, 3]\times[-3, 3]$.  Thus, for the full data problem, we use the true image in the left column of Figure \ref{fig:full_recon}, where every pixel outside of the yellow box is set to zero. Then, for each image, we simulated observations as in \eqref{eq:inverseprob} where we added Gaussian white noise that was scaled so that the noise level was $\frac{\|\bfe\| } {\| \bfA \bff_{\rm true}\|} = 0.02$. 

For this example, we can compute the LS reconstruction using standard techniques (e.g., via the QR factorization \cite{bjorck2024numerical}), and the reconstruction $\widehat \bff$ is provided in the right column of Figure \ref{fig:full_recon}, with the relative reconstruction error, computed as $\frac{\| \widehat \bff - \bff_{\rm true} \|}{\| \bff_{\rm true} \|}$, provided in the title.  We also compute reconstructions obtained using an iterative LS solver, namely LSQR \cite{paige1982lsqr2}.  We observe that after $100$ iterations of LSQR, we get comparably good reconstructions.  However, we do note that for larger noise levels, the LS solution may become contaminated with noise even in the full data case, so regularization may still be required.  For iterative methods such as LSQR, early termination of the iterations is a common approach to avoid potential noise amplification in the solution \cite{hansen2010discrete}. These and other regularization techniques are considered next for the partial data problem.

\subsection{Partial data reconstruction problem}  
Next we consider the partial data reconstruction problem, for which we demonstrate two main points.  First, for the partial data problem, regularization is critical for obtaining a good reconstruction. Second, different regularization choices can be used to incorporate prior knowledge about the solution.  We remark that contrary to our previous CMB reconstruction study \cite{COW}, the unknowns here represent a 2D image (i.e., the initial condition), rather than a 3D spatial-temporal image.  That is, by incorporating the PDE solution operator, we obtain a CMB reconstruction problem that is better conditioned and requires estimating significantly fewer unknown parameters.  Moreover, only spatial regularization is needed here, rather than expensive spatiotemporal priors, which presents a wide array of options for priors.

We consider various forms of regularization. For large-scale problems, iterative methods can be used to compute approximate solutions to \eqref{eq:LS}, where at the $k$th iteration, a solution is sought in a $k$-dimensional Krylov subspace, $\mathcal{K}_k$.  That is, the $k^{\rm th}$ iterate is given by
\begin{equation} \label{hybridTik}
    \bff^{(k)} = \argmin_{\bff \in \mathcal{R}(\mathbf{V}_k)} \left\| \bfA \bff - \bfb \right\|^2
\end{equation}
where $\mathcal{R}(\cdot)$ denotes the range of a given matrix and $\mathcal{R}(\mathbf{V}_k) = \mathcal{K}_k(\mathbf{A}^\top\mathbf{A},\mathbf{A}^\top\mathbf{b})$ when the starting guess, $\mathbf{x}^{(0)}$, is the zero vector.  It is well known that for ill-posed inverse problems, iterative methods such as LSQR exhibit semiconvergence behavior whereby early iterations reconstruction good solution approximations, but latter iterations are dominated by inverted noise \cite{hanke1993regularization}.  An early termination of the iterative method can produce a regularized solution, where the stopping iteration serves as the regularization parameter.

We also consider variational regularization approaches where the goal is to solve a penalized problem of the form,
\begin{equation} \label{penalizedLS}
\min_\bff \| \bfA \bff - \bfb \|^2 + \lambda\,R(\bff)
\end{equation}
where $R$ is a regularization term and $\lambda>0$ is a regularization parameter that balances the trade-off between the data fit and the regularization term. For example, the $\ell_1$ regularization problem,
\begin{equation} \label{l2_l1}
    \min_\bff \left\| \bfA \bff - \bfb \right\|^2 + \lambda\left\|\bff\right\|_1,
\end{equation}
is known to promote sparsity in solutions by reducing the sensitivity to outliers. 
Various algorithms can be used to solve \eqref{l2_l1}, and we use the Fast-ISTA (FISTA) method \cite{FISTA} with backtracking as implemented in the software package IR-Tools \cite{IRtools}.

As an alternative strategy, we consider edge preserving regularization, where an edge preserving prior is used to impose sparsity on derivative images \cite{caselles2015, gonzalez2017}. Following \cite{bardsley2018computational}, we use intrinsic Gaussian Markov random field (IGMRF) priors where the increment variance is allowed to be larger in some locations.  More precisely, recall that the unknown initial state is defined on a two dimensional mesh, so let $f_{ij}$ denote the pixel value at mesh point $(x_i,y_j)$ for $i,j = 1, \ldots, n$. Then, we assume the independent increment model, 
\begin{align*}
\Delta_s f_{ij} & \sim \calN(0, (w_{ij}^s \delta)^{-1}), \quad i = 1, \ldots , n-1, \quad j = 1, \ldots , n\\
\Delta_t f_{ij} & \sim \calN(0, (w_{ij}^t \delta)^{-1}), \quad i = 1, \ldots , n, \quad j = 1, \ldots , n-1
\end{align*}
where the horizontal and vertical increments are given by $\Delta_s f_{ij} = f_{i+1, j} - f_{ij}$ and $\Delta_t f_{ij} = f_{i, j+1} - f_{ij}$ respectively and $\calN$ denotes a normal distribution. If we take $w_{ij} = w_{ij}^s = w_{ij}^t$ and let $\bfLambda = {\rm diag}({\rm vec}(\{w_{ij}\}))$, we get a prior of the form, 
\begin{equation}
\label{eq:edgeprior}
    p(\bff \mid \delta) \propto \delta^{(n^2-2)/2}{\rm exp}\left(-\frac{\delta}{2} \bff\t (\bfD_s\t \bfLambda \bfD_s + \bfD_t\t \bfLambda \bfD_t) \bff\right)
\end{equation}
where $\bfD_s = \bfI \kron \bfD$ and $\bfD_t = \bfD \kron \bfI$ with 
$$\bfD =\begin{bmatrix}
    -1 & 1 & 0 & \cdots & 0 \\
   0  & -1 & 1 & \ddots &\vdots \\
    \vdots & \ddots & \ddots & \ddots & 0\\
    0 &  & \dots & -1 & 1\\
    1 & 0 & \cdots & 0 & -1
\end{bmatrix} \in \bbR^{n \times n}$$
and $\delta >0$.
If the goal is to preserve edges, we can assume that we have an estimate of $\bff$ and take $\bfLambda (\bff) = {\rm diag} ( \bfone ./ \sqrt{(\bfD_s \bff)^2 + (\bfD_t \bff)^2+ \beta \bfone})$ in \eqref{eq:edgeprior} where vector operations are computed element wise and $0<\beta \ll 1$.  We use an iterative process to estimate $\bff$ and use the estimates to update the prior.  See Algorithm \ref{alg:edge} for a description of the edge preserving algorithm.  Note that at each iteration, we must solve a Tikhonov regularization problem, and for this, we use the conjugate gradient (CG) method with a fixed regularization parameter $\lambda_k$. 

\begin{algorithm}
\caption{Edge preserving algorithm using IGMRF}\label{alg:edge}
\begin{algorithmic}
\Require $\bfA$, $\bfb$, $K$ (maximum outer iterations)
\State Set $\bfL_{1} = \bfD_s\t  \bfD_s + \bfD_t\t \bfD_t$
\State Compute $\bff^{(1)} = (\bfA\t \bfA + \lambda_1 \bfL_1)^{-1} \bfA\t \bfb$ using CG
\For{$k = 2,\ldots, K$}
\State Set $\bfL_{k} = \bfD_s\t \bfLambda(\bff^{(k-1)}) \bfD_s + \bfD_t\t \bfLambda (\bff^{(k-1)})\bfD_t$
\State Compute $\bff^{(k)} = (\bfA\t \bfA + \lambda_k \bfL_k)^{-1} \bfA\t \bfb$ using CG
\EndFor
\end{algorithmic}
\end{algorithm}

Next we present various numerical results comparing different reconstruction algorithms. For the numerical experiments presented here, observations have added noise such that the noise level is $0.02$.  We provide the LS solution, the LSQR reconstruction that corresponds to the smallest reconstruction error norm, the FISTA reconstruction after $500$ iterations with $\lambda= 6.6\times 10^{-5}$, and the edge preserving reconstruction with $K=5$, $100$ maximum CG iterations, $\lambda_k = e^k$ where $k$ is the outer iteration counter, and $\beta = 0.001$. Finally, we remark that in the following experiments, the reflected waves from the lateral boundaries are not observed. Thus they do not affect the reconstructions.
 
\begin{figure}[bthp]
\includegraphics[width = 1.1\textwidth]{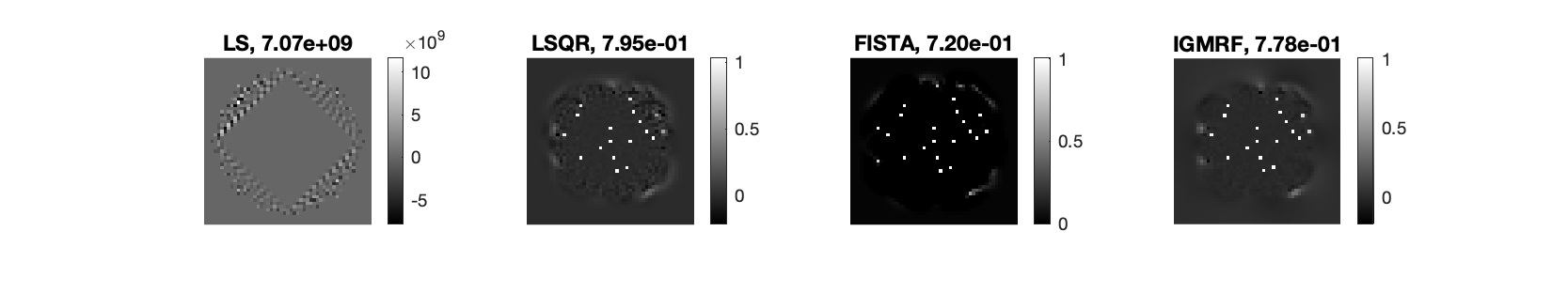}
\includegraphics[width = 1.1\textwidth]{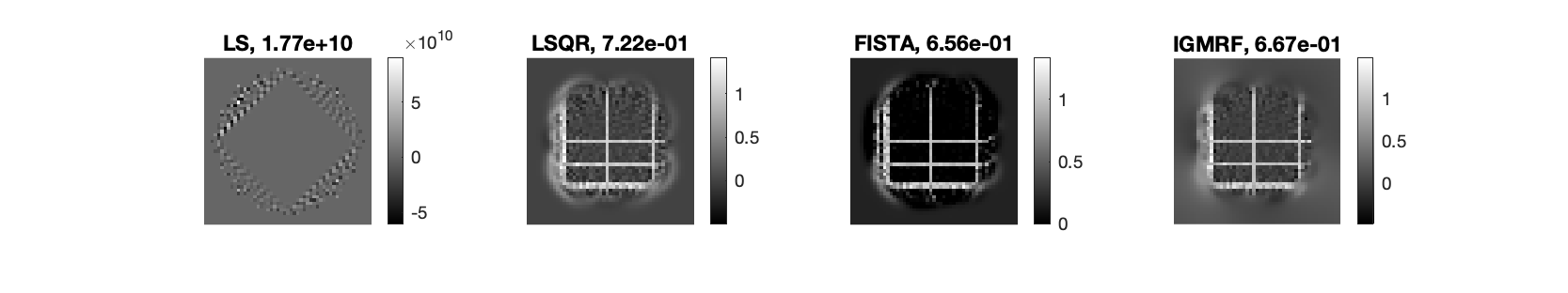}
\caption{Partial data reconstructions for the $7 \times 7$ detector grid for the random dots example (top) and the random lines example (bottom). Relative reconstruction error norms are provided in the titles.}
\label{fig:partial_recon_7}
\end{figure}

\begin{figure}[bthp]
\includegraphics[width = \textwidth]{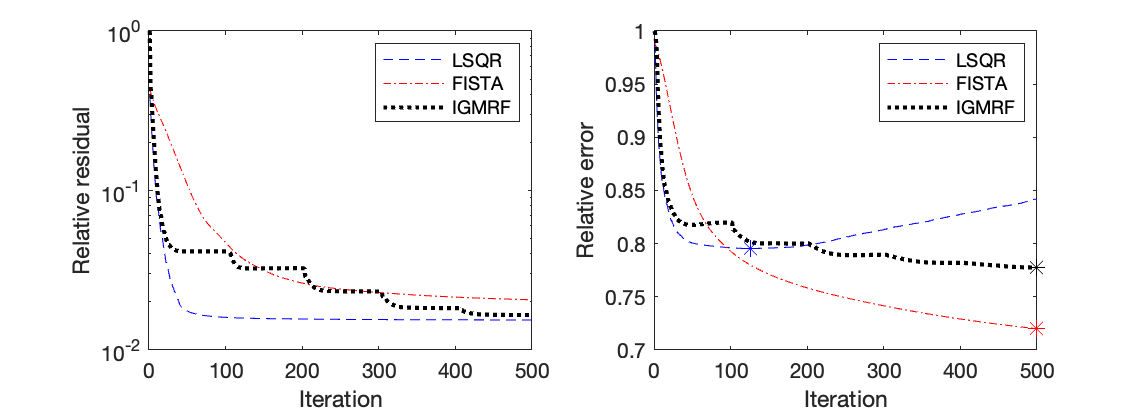}
\includegraphics[width = \textwidth]{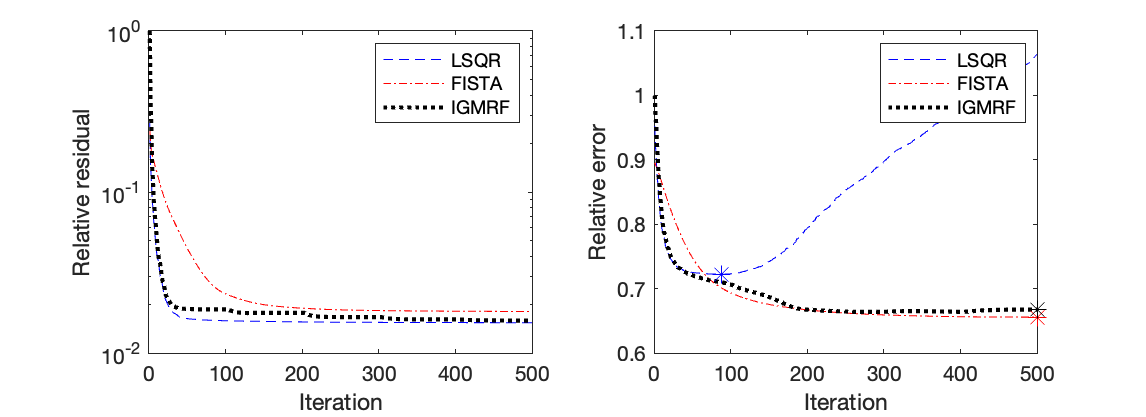}
\caption{Relative residual norms (left) and relative reconstruction error norms (right) per iteration for the partial data reconstruction problems.  These results correspond to the $7 \times 7$ detector grid for the random dots example (top) and random lines example (bottom).}
\label{fig:partial_res}
\end{figure}

First, for the tomography reconstruction problem with a $7 \times 7$ detector grid, we provide reconstructions for the random dots example and for the random lines example in Figure \ref{fig:partial_recon_7}. In Figure \ref{fig:partial_res}, we provide the relative residual norm per iteration and the relative reconstruction error norms per iteration.  We observe that without regularization, the least-squares reconstruction is very poor because the reconstruction is dominated by error and noise. The LSQR reconstruction is significantly better and corresponds to the LSQR iterate with the smallest relative error norm. This is marked by a blue star in the right figures of Figure \ref{fig:partial_res}. Although the error curves are not available in practice, the results show that if a good stopping iteration or stopping criteria is known, then the LSQR reconstruction is good. However, without an appropriate stopping criteria, the reconstruction can become dominated with noise.  From the plot of the relative errors for LSQR, we observe semiconvergence whereby the errors decrease in early iterations but increase at later iterations. 

Both FISTA and the edge-preserving algorithm provide good reconstructions. 
Notice that since IGMRF has an inner and outer iteration, the x-axis in Figure \ref{fig:partial_res} for IGMRF corresponds to the total number of CG iterations.  Moreover, the stars correspond to the reconstructions in Figure \ref{fig:partial_recon_7}. We note that due to the limited view from the detectors, only unknowns in the central region of the image are able to be reconstructed accurately. Regions along the borders are not visible and hence do not get reconstructed.

\begin{figure}[bthp]
\includegraphics[width = \textwidth]{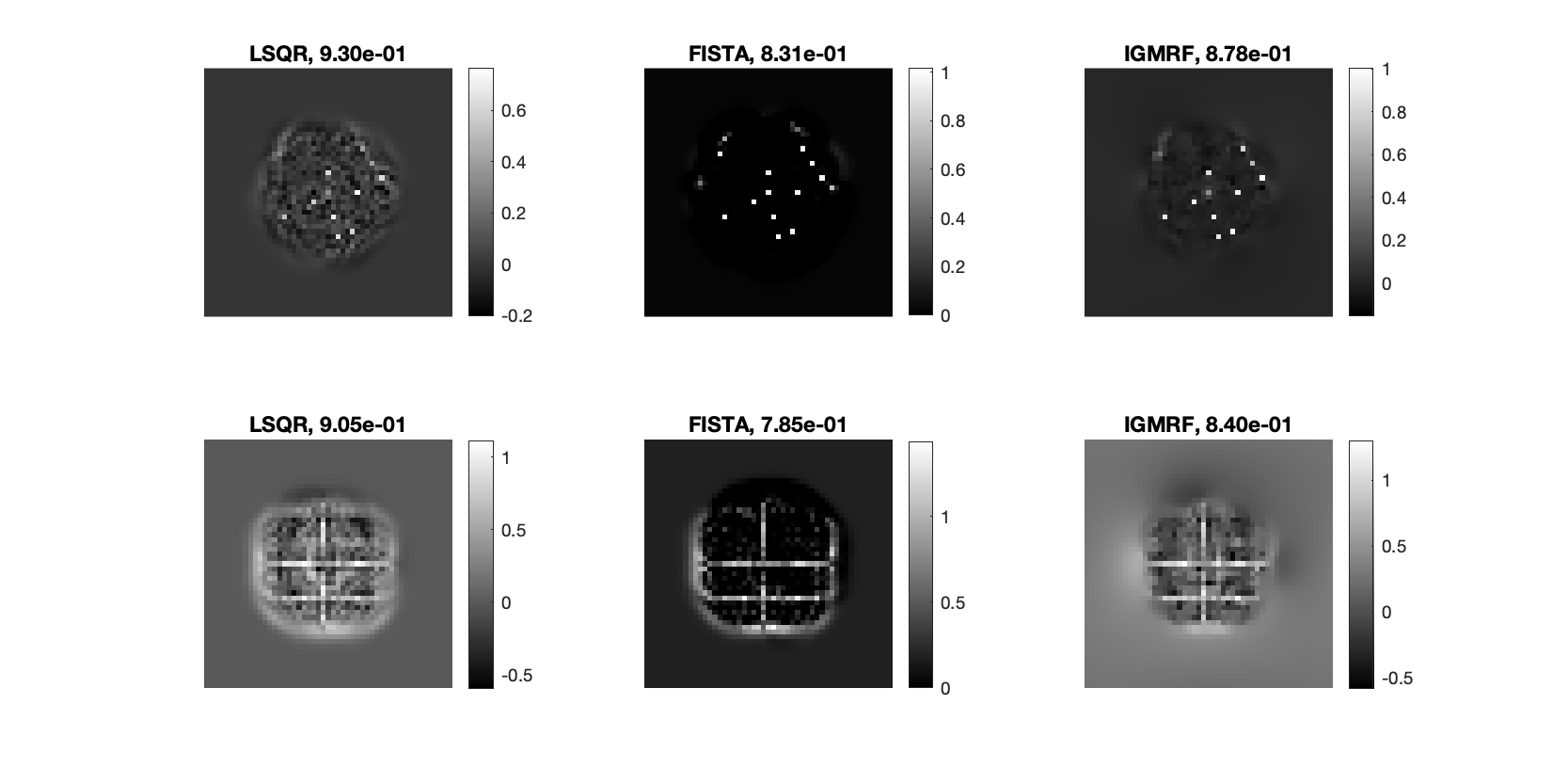}
\caption{Reconstructions for LSQR, FISTA, and IGMRF for the partial data problem with a $3 \times 3$ detector grid.  The top row contains reconstructions for the random dots example and the bottom row contains reconstructions for the random lines example. Relative reconstruction error norms are provided in the titles.}
\label{fig:partial_3}
\end{figure}

For a smaller $3 \times 3$ detector grid, we compare reconstructions for LSQR, FISTA, and IGMRF in in Figure \ref{fig:partial_3}.  We do not provide results for the LS solution because due to the ill-posed nature of the problem, the LS solution is useless (i.e., dominated by noise as illustrated in the leftmost column of Figure \ref{fig:partial_recon_7}). In both examples, we observe that due to the smaller grid of detectors, the window or region of reconstruction is smaller.  
In particular, there is a very small circular region in the center of the image that cannot be reconstructed as well as the ring region. This agrees with our theoretical analysis. However, we observe that the size of the ring region seem to be larger than that predicted in Theorem \ref{thm-main}. This is not an issue because our initial data is not supported in the ring region as required by Theorem \ref{thm-main}, and in addition we include regularization. Nevertheless, the numerical results seem to indicate that at least some partial information can be reconstructed in a larger region inside the ring, which agrees with our remark in the end of Section \ref{sec-pf1}.

\section{Conclusions}
\label{sec-conc}

In this work, we have shown that it is possible to recover the initial status of the observable universe using only the  CMB observed near the Earth. Mathematically, we show that the inversion of the cosmological X-ray transform, which is a severely ill-posed inverse problem, can be stabilized by incorporating the physical model. In addition to the analysis, we provide a numerical study for inverting the ray transform. We investigate the spectral properties of the discretized problem which reveal the importance of incorporating regularization for the partial data problem. Also, we compare various reconstruction methods to demonstrate the feasibility of the tomography method for cosmological applications.

Our study opens up many possibilities. One outstanding problem in cosmology is the detection of primordial gravitational waves generated in the beginning of Universe, see for example \cite{KDS}. Theoretical studies have shown that these waves will leave signatures (e.g., polarizations) in the CMB, although such signatures have not been identified yet. The tomography approach we studied in this paper provides another way to find the primordial waves indirectly. For this purpose, one needs to study the inversion of the cosmological X-ray transform acting on metric tensors. This is a more challenging problem due to the presence of a non-trivial null space for the transform, see \cite{LOSU, Wan2}. Another interesting problem is to develop accurate and efficient numerical methods for simulations on a larger scale, especially for the realistic 3D universe. In particular, the tomography method can be adapted to geometric backgrounds (instead of the flat Minkowski background) and handle CMB data collected along trajectories of the satellites. 

\section*{Acknowledgments} 
This work was partially supported by the National Science Foundation (NSF) under grants DMS-2411197 (J. Chung) and DMS-2205266 (Y. Wang).

\bibliographystyle{abbrv} 
\bibliography{references} 
\end{document}